\documentclass[aps,pra,floatfix,twocolumn,showpacs,tightenlines,groupedaddress,superscriptaddress,amsmath]{revtex4}
\usepackage{amstext}
\usepackage{bbm}
\usepackage{amssymb}
\usepackage{amsmath}
\usepackage{amsfonts}
\usepackage{mathbbol}
\usepackage{placeins}
\usepackage{comment}
\usepackage{dsfont}
 \usepackage[pdftex]{color,graphicx}
\newcommand{\be}{\begin{equation}}
\newcommand{\ee}{\end{equation}}
\newcommand{\ba}{\begin{eqnarray}}
\newcommand{\ea}{\end{eqnarray}}
\newcommand{\ha}{\hat{a}}
\newcommand{\hadag}{\hat{a}^\dagger}

\begin{document}

\title{Continuous Measurement of an Atomic Current}
\author{C. Laflamme}
\author{D. Yang}
\author{P. Zoller}
\address{Institute for Theoretical Physics, University of Innsbruck, A-6020 Innsbruck, Austria}
\address{Institute for Quantum Optics and Quantum Information of the Austrian Academy of Sciences, A-6020 Innsbruck, Austria}

\date{\today}
\begin{abstract}
We are interested in dynamics of quantum many-body systems under continuous observation, and its physical realizations involving cold atoms in lattices. In the present work we focus on continuous measurement of atomic currents in lattice models, including the Hubbard model. We describe a Cavity QED setup, where measurement of a homodyne current provides a faithful representation of the atomic current as a function of time. We employ the quantum optical description in terms of a diffusive stochastic Schr\"odinger equation to follow the time evolution of the atomic system conditional to observing a given homodyne current trajectory, thus accounting for the competition between the Hamiltonian evolution and measurement back-action. As an illustration, we discuss minimal models of atomic dynamics and continuous current measurement on rings with synthetic gauge fields, involving both real space and synthetic dimension lattices (represented by internal atomic states). Finally, by `not reading' the current measurements the time evolution of the atomic system is governed by a master equation, where - depending on the microscopic details of our CQED setups - we effectively engineer a current coupling of our system to a quantum reservoir. This provides novel scenarios of dissipative dynamics generating `dark' pure quantum many-body states.
\end{abstract}
\pacs{}

\keywords{}
\maketitle

\section{Introduction}

Quantum optics defines and implements the paradigm of continuous measurement
of a quantum system, where the system of interest is monitored continuously
by measuring the scattered light in a photon counting or homodyne
experiment. Given a sequence of photon counts, or homodyne current trajectory, continuous measurement theory provides the description
of the associated conditional time evolution of the system \cite{QuantumNoise,Molmer:1991aa, Carmicheal:1993aa, Barchielli1991,Wiseman1993,Diosi:1988aa}.
This evolution is written in terms of a Stochastic Schr\"odinger Equation
(SSE) for a wave function $|\psi_{c}(t)\rangle$, where the time evolution
of the system is a competition between dynamics induced by the system
Hamiltonian and the measurement back-action on the system. Motivated
by recent advances in building controlled quantum many-body systems
with quantum optics, and, in particular, in realizing strongly correlated
atomic dynamics and quantum phases such as Hubbard models \cite{Greif2016:aa, Mancini:2015aa, Stuhl:2015aa, Jotzu:2014aa, Aidelsburger:2013aa,Struck:2013aa,Bakr2009:aa, Sherson2010:aa, Cheuk2015:aa, Haller2016:aa}, below we are
interested in developing and applying the theory of \emph{continuous observation}
to \emph{quantum many-body systems.} This requires first of all the identification
of physical observables of interest to be monitored in time,
and, in particular, the design and description of the underlying measurement
setups within microscopic quantum optical models compatible with
the requirements of (non-destructive) continuous observation. We note that such measurements
scenarios can be both quantum demolition or quantum non-demolition
in nature. 
\begin{figure}[t!]
\centering{}\includegraphics[width=0.45\textwidth]{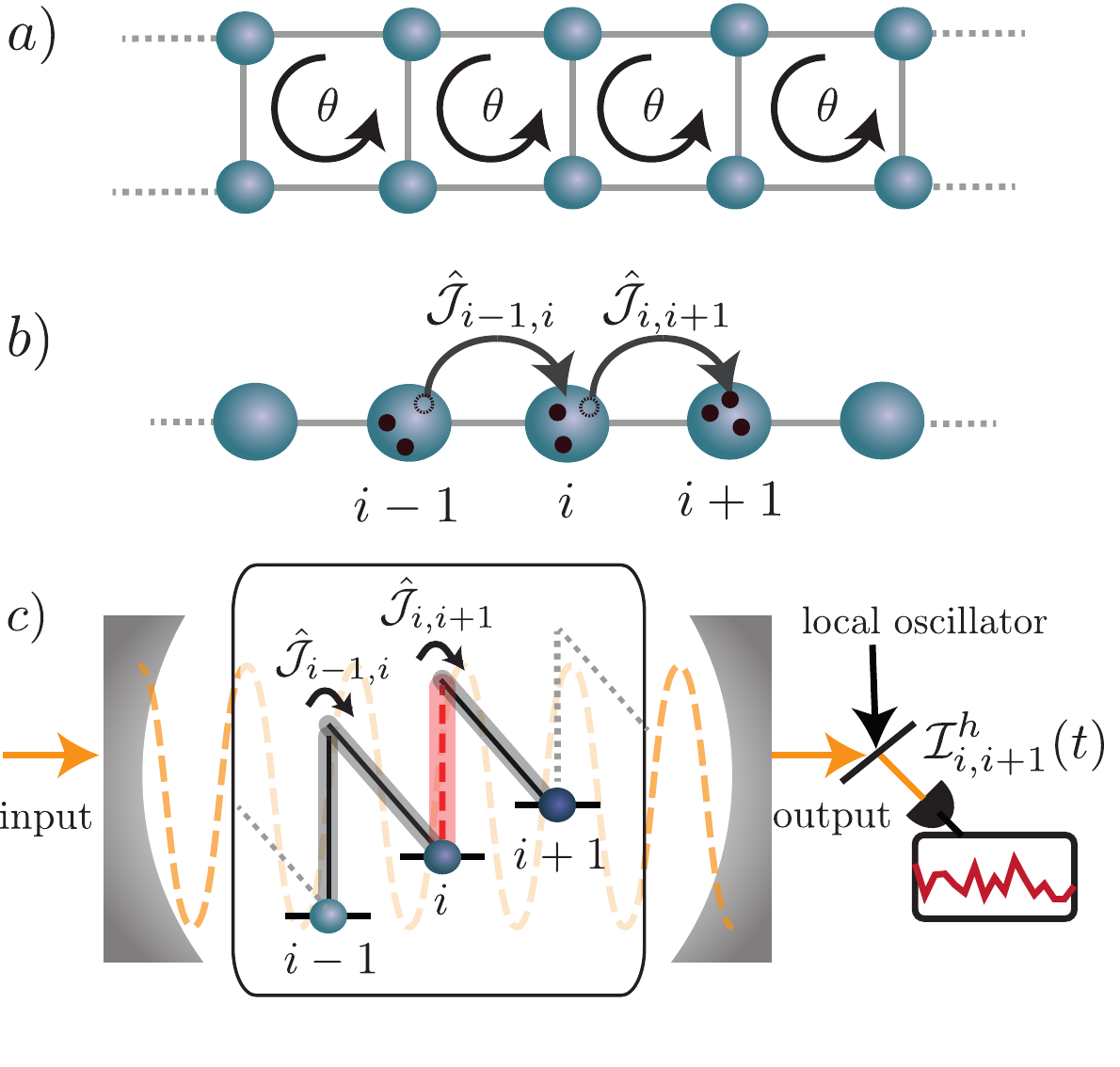} \caption{  Schematic of the current operators $\mathcal{J}_{i,i+1}$ and measurement setup in a cold atom context. a) A 2D lattice
system with finite flux $\theta$ resulting in a persistent current
on each plaquette, and b) a 1D setup. The current operator $\mathcal{J}_{i,i+1}$ involves the particles moving from site $i$ to site
$i+1$. c) Measurement schematic: Transitions between sites are made via Raman transitions driven by external lasers (black/solid), and those to be measured are additionally coupled to a cavity field (red/dashed). The cavity output is mixed with a local oscillator resulting
in a homodyne current $\mathcal{I}^h(t)$ proportional to the local current.
Coupling to all transitions results in a measurement of global current
$\hat{\mathcal{J}}_{\textrm{tot}}$.}
\label{fig:schematic1} 
\end{figure}

In the present paper we focus on \emph{continuous measurement of an
atomic current} in lattice models for atomic quantum
many-body dynamics including the Hubbard model. Motivation for the
measurement comes from the ongoing discussion of transport in atomic
circuits coupled to particle reservoirs \cite{Chien:2015aa, Eisert:2015aa, Brantut:2012aa, Schneider:2012aa}, but also from
Hubbard physics with artificial magnetic fields (synthetic gauge fields) \cite{Jaksch03, Zoller16:aa, Boada2012:aa,Celi2014:aa},
where current patterns distinguish various quantum phases \cite{Wen:2004aa}. The kind
of Hubbard models and current measurements we have in mind are depicted
in Fig.~\ref{fig:schematic1}a) and b) for an optical lattice setup. For example, in 1D (c.f.
Fig~\ref{fig:schematic1}b) a relevant Hamiltonian will involve a kinetic energy term
of the form~\cite{Zoller16:aa}

\begin{equation}
H_{J}=-J\sum_{j}e^{i\theta}\hat{a}_{j}^{\dagger}\hat{a}_{j+1}+{\rm h.c.},
\label{eq:HamJ_intro}
\end{equation}
where $\hat{a}_{j}$ ($\hat{a}_{j}^{\dagger}$) are annihilation (creation)
operators for bosonic (or fermionic) atoms at site $j$ and the phase
$\theta$ represents a gauge field in the spirit of a Peierls substitution \cite{Peierls}.
The atomic current as the observable of interest can either be a \emph{local}
current
\begin{equation}
\label{eq:current_example}
\hat{\mathcal{J}}_{j,j+1}=-J\left(ie^{i\theta}\hat{a}_{j}^{\dagger}\hat{a}_{j+1}-ie^{-i\theta}\hat{a}_{j+1}^{\dagger}\hat{a}_{j}\right).
\end{equation}
describing the transfer of particles between two adjacent lattice
sites $j,j+1$ (c.f. Fig~\ref{fig:schematic1}b)), or a \emph{global} current $\hat{\mathcal{J}}_{\textrm{tot}}=\sum_{j}\hat{\mathcal{J}}_{j,j+1}$,
where we integrate over a spatial region. Our aim is to devise a continuous
measurement scheme where, in a given run of the experiment, the atomic
current between two lattice sites is mapped to fluctuations of a laser
beam. These fluctuations can be detected in a homodyne measurement, so that - up
to shot noise - the \emph{homodyne current }measured at a photodiode
is a faithful representation of the \emph{atomic current}, 
\begin{equation}
\label{eq:homodyne_Current}
\mathcal{I}_{j,j+1}^{h}(t)=\sqrt{\gamma}\langle\psi_{c}(t)|\hat{\mathcal{J}}_{j,j+1}|\psi_{c}(t)\rangle+\xi_{j,j+1}(t).
\end{equation}
Here $\mathcal{I}_{j,j+1}^{h}(t)$ is the homodyne current at time
$t$, which follows the expectation value of the current operator
in state $|\psi_{c}(t)\rangle$ with $\gamma$ a measurement strength
and shot noise $\xi_{j,j+1}(t)$. While a scheme to destructively measure the local 
current statistics between two sites was given in \cite{Kesler:2014aa}, we emphasize that we focus here
on building a scheme for the continuous measurement of local and global currents. 

\begin{figure}[t!]
\centering{}\includegraphics[width=0.45\textwidth]{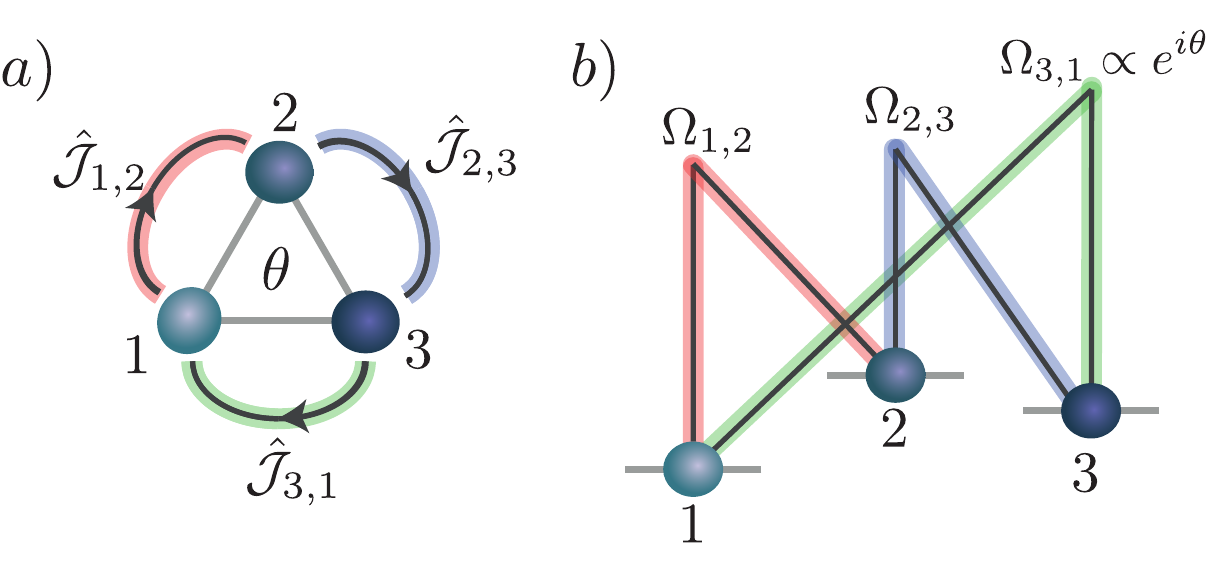} \caption{Current operators, for a 3-site (state) system with enclosed flux $\theta$ (atomic version of a SQUID) realized equivalently with a) external
degrees of freedom on an optical lattice, and b) internal atomic degrees of
freedom, in a trapped gas. }
\label{fig:schematic2} 
\end{figure}

The quantum optical setup we propose, which results in the mapping of an atomic current
to a homodyne current given in Eq.~\eqref{eq:homodyne_Current}, builds on recent
achievements in combining quantum degenerate gases and atomic quantum
simulation with optical Cavity QED (CQED) \footnote{ While there has been much work done on atoms coupled to CQED, as well as on nanostructures \cite{Klinder2015:aa,Douglas:2015aa, Sames:2014aa, Ritsch:2013aa,Thompson2013:aa,Clark:2003aa,Chang2012:aa, Klinder2015:bb}, we emphasize our goal here is to use the cavity solely as a measurement tool, converting the atomic current into a measurable signal.} The scheme we have in mind is for light assisted hopping via Raman induced
coupling, involving a laser driven quantized cavity mode, as illustrated in Fig.~\ref{fig:schematic1}c). In transferring an atom between two
sites, involving the cavity field as one Raman leg, a photon will be
\emph{absorbed} into the cavity field for the transition from site
$j$ to $j+1$, while a photon will be \emph{emitted} in a hopping
process $j+1$ to $j$ from the coherent drive. It is this emission/absorption, i.e. the back-action
on the cavity field with the transfer of atoms, which we wish to monitor
with homodyne detection of light transmitted through the cavity. As Raman induced tunnelling can be implemented for both external (lattice) transitions and internal state transitions, our discussion is applicable to both scenarios (c.f. Fig.~\ref{fig:schematic2}). Finally, we note that the idea of coupling a cavity to a Raman transition has been applied in the case of fermionic systems, for implementation of the Dicke model, and resulting in induced chiral states \cite{Zheng:2016aa, Kollath:2016aa, Sheikhan:2016aa, Baden:2014aa}. By contrast, here the goal of this coupling is as a measurement tool, converting the lattice current into a measurable signal.   

%

The time evolution of our atomic many-body system, subject to the continuous
measurement of the current described in the previous paragraph,
will reflect both evolution according to some Hubbard Hamiltonian $H$, and the back-action from the measurement. 
For a single run of the experiment, the time evolution conditioned on the observed homodyne current $\mathcal{I}_{j,j+1}^{h}(t)$ is described by the wave function $|\psi_{c}(t)\rangle$ obeying the diffusive SSE (for a detailed derivation and references see ~\cite{QuantumWorld})
\begin{equation}
d|\psi_{c}(t)\rangle=\left(-i\hat{H}_{{\rm eff}}\;dt+\sqrt{\gamma}e^{i\varphi}\hat{c} \,dq(t) \right) |\psi_{c}(t)\rangle.\label{eq:homodyneSSE-1}
\end{equation}
Here $\hat{H}_{{\rm eff}}=H-i\gamma\hat{c}^\dagger\hat{c}$ is the (non-Hermitian) effective Hamiltonian, $dq(t)$ is the quadrature component (defined by phase $\varphi$) read in a particular run of the experiment from the output field of the cavity in a time interval $[t,t+dt)$ in homodyne detection and we identify  $dq/dt \equiv \mathcal{I}^{h}(t)$ with the homodyne current of Eq.~(\ref{eq:homodyne_Current}). The atomic operator $\hat{c}$ describes the measurement back-action on the atomic wavefunction, which we will derive below for specific CQED measurement setups. Thus Eqs.~(\ref{eq:homodyne_Current}) and (\ref{eq:homodyneSSE-1}) provide the desired description of homodyne measurement, with the homodyne current following the atomic current in the spirit of continuous observation, and the corresponding back-action on the atomic evolution.

Our discussion so far has focused on continuous measurement of the atomic current via homodyne detection in a given experimental run of the experiment, where the time evolution of the observed systems is described  by the SSE in Eq.~(\ref{eq:homodyneSSE-1}). Instead we can also decide {\em not to read the measurement}, so that the time evolution of the system traced over all possible measurement outcomes is described by a master equation for a system density operator $\rho(t)$ derived from the SSE. We note that this master equation describes the time evolution of a quantum many-body system, which is coupled to a reservoir according to our current measurement, i.e.~a current coupling. This can be understood as {\em quantum reservoir engineering with current coupling} to an environment.  Below we will derive this master equation for various microscopic models of our CQED measurement scenarios, and thus system-environment couplings.  Interestingly we find that for certain combinations of atomic Hamiltonian and parameters, and reservoir coupling these master equations possess steady states in form of pure dark states, $\rho (t) \rightarrow |\Psi \rangle\langle \Psi|$, i.e.~we can drive the open quantum many-body system into nontrivial quantum phases \cite{Poyatos:1996aa}.


The paper is organized as follows: We begin by deriving in detail the theory behind the proposed continuous measurement, giving two possible microscopic implementations through which it can be realized. Both implementations result in a measurement with homodyne current of the form Eq.~\eqref{eq:homodyne_Current}, but differ in the back-action in the associated SSE (c.f. Eq.~\eqref{eq:homodyneSSE-1}). Following the derivation, in Sec.~\ref{sec:BHmodel} we solve the dynamics of an atomic Bose-Hubbard system under the continuous measurement of current. We will consider first a simple illustration in a simple 3-level-system that can be reduced to a double well, with macroscopic tunnelling between the two ground states, which have equal and opposite current. After gaining intuition through this example, we will integrate the SSE for several different example model systems, highlighting the impact of the associated back-action in the case of both QND and non QND measurement. Finally we will briefly describe the evolution of the density matrix with the associated master equation.

\section{Continuous Measurement of an Atomic Current with a CQED setup }
\label{sec:measurement_setup}
In this section we propose two different microscopic schemes to couple the atomic system of interest to the cavity mode, deriving the corresponding equations of motion. We will show that, while both schemes result in the desired homodyne current (c.f. Eq.~\eqref{eq:homodyne_Current}), the difference in microscopic details result in a different back-action response of the system, which is reflected in differences in the associated SSE. To concentrate on the details of the derivation, we focus here on an atomic system containing just two states, which is the basic building block of our proposal; in further sections we will be able to take this building block and expand on it into a physically relevant system. 

\begin{figure}[t!]
\begin{center}
	\includegraphics[width=0.45\textwidth]{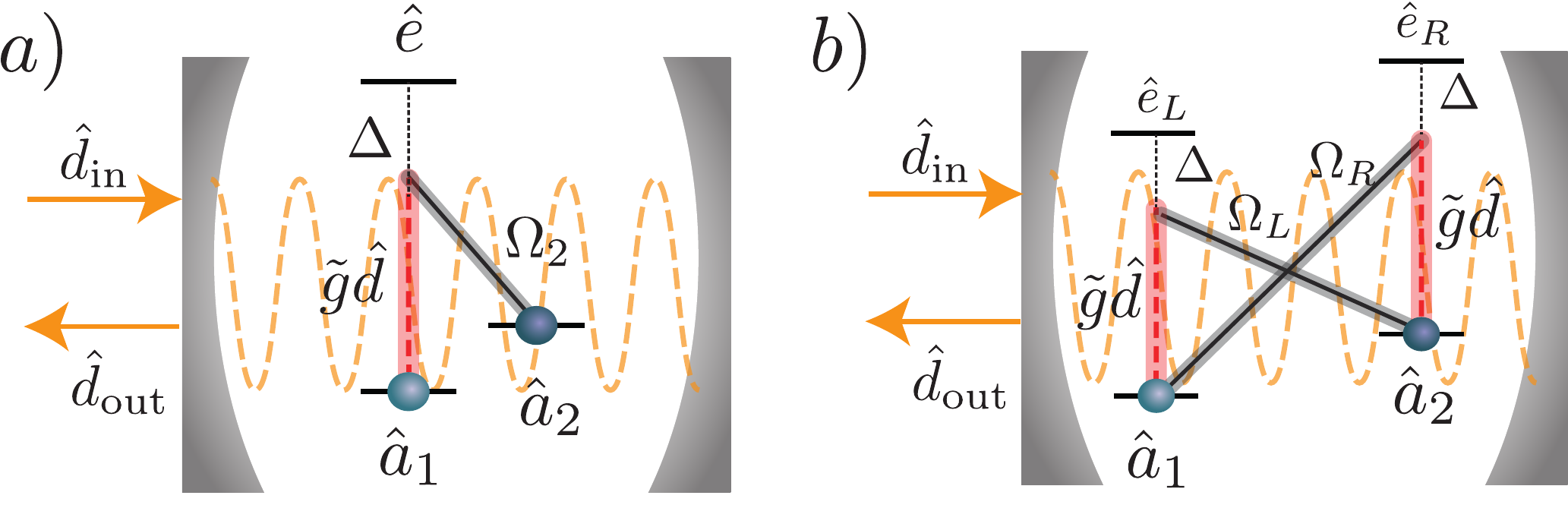}
	\caption{Schematic of the atom-cavity coupling for a) the asymmetric coupling and b) the symmetric coupling. In both cases, the Raman transition mitigates hopping between the system levels $\hat{a}_{1/2}$ where one, or both legs of the Raman transition are coupled via the bare coupling $\tilde{g}$ to a cavity mode $\hat{d}$. Together with standard laser fields with Rabi frequencies $\Omega_2$ and $\Omega_{L/R}$ the effective couplings are given, for large detunings, by $g=\tilde{g}\Omega_2/\Delta$ and $g_{L/R}=\tilde{g}\Omega_{L/R}/\Delta$  in the asymmetric and symmetric systems, respectively. The cavity is driven with an input beam $\hat{d}_{\rm in}$, which is assumed to be the vacuum field, and the output $\hat{d}_{\rm out}$ is measured via homodyne measurement.  }
	\label{fig:schematic}
\end{center}
\end{figure}

In the derivation below we begin with the first of our two proposed microscopic coupling schemes, shown schematically in Fig.~\ref{fig:schematic}a), deriving all expressions for this case. In the subsequent section we will then introduce the second scheme, shown schematically in Fig.~\ref{fig:schematic}b), and, because the derivation is completely analogous, just highlight the resulting equations. 
\subsection{Asymmetric Raman Coupling to the Cavity}
\subsubsection{Hamiltonian}
We begin with the total Hamiltonian for our first coupling scheme, shown schematically in Fig.~\ref{fig:schematic} a). The Hamiltonian governing the complete system is given by 
\be
\hat{H}=\hat{H}_{\rm sys}+\hat{H}_{\rm bath}+\hat{H}_{\rm int},
\ee
where $\hat{H}_{\rm sys}$ is the Hamiltonian governing the atomic system, the cavity, and the coupling between them, $\hat{H}_{\rm bath}$ is the bath which is coupled to the cavity allowing us to measure the output, and $\hat{H}_{\rm int}$ is the coupling between the cavity and this bath. 
As mentioned in the introduction, we propose two different microscopic schemes in which the atomic system and the cavity are coupled, thus the two schemes will involve a different coupling term within $\hat{H}_{\rm sys}$. The first of which is ($\hbar=1$ in all of what follows)
\ba
\label{eq:coupling_asym}
\hat{H}_{\rm sys}&=& \omega_{\rm c} \hat{d}^\dagger \hat{d}+\omega_{21}\hat{a}_2^\dagger \hat{a}_2+g \hat{a}_1^\dagger \hat{a}_2 \hat{d}^\dagger+ g^* \hat{d} \hat{a}_2^\dagger \hat{a}_1 + \hat{H}_{\rm bh}.
\ea
Here, $\hat d$ is the destruction operator of the cavity mode, which has frequency $\omega_{\rm c}$, and $\hat{a}_1,\hat{a}_2$ are the destruction operators of the two (bosonic) internal states, with energy difference $\omega_{21}$. 
The Hamiltonian governing the desired dynamics of the atomic system is $\hat{H}_{\rm bh}$. In later sections we will take this to be the Bose-Hubbard Hamiltonian, however the exact form of this Hamiltonian plays no role in the derivation here. 
Finally, the coupling $g\in \mathbb{C}$ between the bosonic system and cavity is an effective coupling obtained via a Raman transition, where the intermediate (excited) state manifold has been eliminated at large detuning (see Fig.~\ref{fig:schematic}). The coupling scheme in Eq.~\eqref{eq:coupling_asym} will be called the {\it asymmetric} scheme, as the coupling implies a photon is added to the cavity when an atom transfers from system site $\hat{a}_1$ to $\hat{a}_2$ but annihilated when an atom is transferred from $\hat{a}_2$ to $\hat{a}_1$.

Regardless of the scheme, the Hamiltonian of the external bath is modelled by
\ba
\hat{H}_{\rm bath}&=& \int d\omega \; \omega \; \hat{b}^\dagger(\omega)\hat{b}(\omega),
\ea
where the integral over $\omega$ is assumed to be over the relevant bandwidth of the bath. Finally, the coupling between the external bath and cavity is
\ba
\hat{H}_{\rm int}&=& i\sqrt{\frac{\kappa}{2\pi}} \int d\omega \left(\hat{b}^\dagger(\omega) \hat{d} - \hat{d}^\dagger \hat{b}(\omega)\right) ,
\ea
where we have assumed that the Markov approximation holds, such that the coupling of the cavity to the bath is constant over the bandwidth, and given by $\sqrt{\kappa/2\pi}$.

There are two ways in which one can view the role of the external field: the first, is to assume the field is in the vacuum state, where only vacuum fluctuations are at the input of the cavity. In this case, the role of the field is solely to make the measurement, and the transitions between states (which will be part of the Hamiltonian term $H_{\rm bh}$) must be induced with an independent laser.  The second case is to assume the field is in a coherent state, where the cavity's average field will induce these transitions. Via a Mollow transformation \cite{Mollow:19}, these two scenarios are mathematically equivalent, and for simplicity we will consider the former. 
 
Note that, for clarity of the following discussion, we have neglected any Stark shifts on the states  $1$ and $2$, and we neglect spontaneous emission from the excited manifold in the limit of large detuning; these effects can, however, be readily added to the present model.

\subsubsection{Derivation of the SSE and Homodyne Current}
We now derive the time evolution due to the Hamiltonian presented in the previous subsection; our derivations are based on those carried out in \cite{QuantumNoise}. 
We begin by moving into the interaction picture, transforming away the optical frequencies in $\hat{H}_{\rm sys}$ and in the free Hamiltonian $\hat{H}_{\rm bath}$. This transformation results in time-dependent operators $\hat{d}\rightarrow e^{-i\omega_{\rm c}t}\hat{d}, \hat{a}_{2}\rightarrow e^{-i\omega_{21}t}\hat{a}_2$ as well as the bath modes, for which we define the time-dependent operator 
\be
\hat{b}(t) =\frac{1}{\sqrt{2\pi}} \int d\omega e^{-i\omega t} \hat{b}(\omega), \qquad [\hat{b}(t),b^\dagger(t')] = \delta (t-t').
\ee
For clarity, we will drop the time argument of these operators from this point, however we emphasize we are in the interaction picture where the Hamiltonian changes accordingly, and part of the time dependence is in the operators themselves. In this interaction picture, the time evolution operator $\hat{U}(t)$ obeys the Schr\"odinger equation,
$
\frac{d}{dt}\hat{U}(t) = -i \hat{H}\hat{U}(t)$ with $ \hat{U}(0)=1
$.
This equation can be interpreted as a Stratonovich $(S)$ quantum stochastic Schr\"odinger equation (QSSE),  
\ba
(S) \;d\hat{U}(t) &=& \bigg[-i \left(\hat{H}_{\rm bh}+ g \hat{a}_1^\dagger \hat{a}_2 d + g^* \hat{d}^\dagger \hat{a}_2^\dagger \hat{a}_1\right)dt \nonumber \\&&+ \sqrt{\kappa}\left( d\hat{B}^\dagger \hat{d} - \hat{d}^\dagger d\hat{B}\right) \bigg]\hat{U}(t),
\ea
where we have defined stochastic increments
\be
d\hat{B} = \int_t^{t+dt}dt'\, b(t').
\ee
While the Stratonovich form allows us to relate to the physical system, we must transform into Ito form  $(I)$ in order to take advantage of Ito calculus. After this transformation we obtain
\ba
\label{eq:Ito_U_eqn}
(I) \;d\hat{U}(t) &=& \bigg[ \left(-i\hat{H}_{\rm bh}-i g \hat{a}_1^\dagger \hat{a}_2 d -i g^* \hat{d}^\dagger \hat{a}_2^\dagger \hat{a}_1 - \frac{\kappa}{2} \hat{d}^\dagger d\right)dt \nonumber \\ &&+\sqrt{\kappa}\left( d\hat{B}^\dagger \hat{d} - \hat{d}^\dagger d\hat{B}\right) \bigg]\hat{U}(t).
\ea
We consider the case where the external field is in the vacuum state, where $\hat{b}|{\rm vac}\rangle =0$,  and the Ito increments $d\hat{B}$ satisfy the standard Ito rules
\ba
d\hat{B}^\dagger d\hat{B}&=&0 = d\hat{B}^\dagger d\hat{B}^\dagger=d\hat{B}d\hat{B}\nonumber \\
d\hat{B}d\hat{B}^\dagger&=&dt.
\ea
We are now prepared to write the time evolution of the state vector $|\Psi(t)\rangle = \hat{U}(t)|\Psi(0)\rangle$, which, from Eq.~\eqref{eq:Ito_U_eqn} is given by the QSSE
\ba
\label{eq:Ito_QSSE_orig}
(I) \;d|\Psi(t)\rangle &=& \bigg[ \left(-i\hat{H}_{\rm bh}-ig \hat{a}_1^\dagger \hat{a}_2 \,\hat{d}^\dagger -i g^* \hat{d}\, \hat{a}_2^\dagger \hat{a}_1  \right)dt\nonumber \\ &&- \frac{\kappa}{2}\left( \hat{d}^\dagger  \hat{d} \right) dt+ \sqrt{\kappa}\left( d\hat{B}^\dagger\hat{d}\right) \bigg] |\Psi(t)\rangle
\ea
Physically, Eq.~\eqref{eq:Ito_QSSE_orig} includes the following components: evolution of the system with the desired Hamiltonian (in this case $H_{\rm bh}$) and the coupling of the system to cavity, which results in photons added or subtracted from the cavity when a transition between states occurs. The final terms describe the stochastic evolution due to the coupling of the cavity to an external bath, which allows us to measure the light leaving the cavity. 

Rather than measuring the amplitude of this outgoing light directly, we focus on making a homodyne measurement, which gives access to quadrature (i.e. phase) information. As we will show, this quadrature information will be necessary to relate the homodyne current to the atomic current that we are after. Anticipating this measurement, we define a quadrature of the Ito increment
\be
\sqrt{2}d\hat{Q} =d\hat{B}e^{i\varphi}+d\hat{B}^\dagger e^{-i\varphi},
\ee
where $\varphi$ is a phase whose value will be specified in a later section. Given that $d\hat{B}$ acting on the vacuum gives a zero term, the QSSE in Eq.~\eqref{eq:Ito_QSSE_orig} can be equivalently rewritten as 
\ba
\label{eq:Ito_QSSE}
(I) \;d|\Psi(t)\rangle &=& \bigg[ -i \left(\hat{H}_{\rm bh}+g \hat{a}_1^\dagger \hat{a}_2 \hat{d}^\dagger +g^* \hat{d} \hat{a}_2^\dagger \hat{a}_1\right)dt\nonumber \\ && - \frac{\kappa}{2} \big(\hat{d}^\dagger d\big)+ \sqrt{2\kappa} e^{i\varphi}d\hat{Q}\hat{d} \bigg] |\Psi(t)\rangle
\ea
As an equation for the total state vector $|\Psi(t)\rangle$, the QSSE in Eq.~\eqref{eq:Ito_QSSE} represents the evolution of the system plus external electromagnetic field. However, here we are interested in what happens to the system {\it conditioned} on a specific sequence of measurements of the electromagnetic field, with one measurement made during each time interval $dt$. If the magnitude of the external field was measured directly, one would obtain zero if, during the time interval $dt$, no photon was recorded, or one, if a photon was recorded. In contrast, here we want to make a measurement of the quadrature of the outgoing light. In practice this is done by mixing the output light of the cavity with a classical reference beam representing a local oscillator and measuring the resulting combined field. Because we are assuming the input field is in a vacuum state, this classical beam would come from an independent source. In the case of a coherent input field, this input field can itself be used as the classical beam. Regardless of the source of the classical field, the total measured signal takes the form
\ba
\mathcal{I}^h(t) &=& \langle (\beta +\hat{b}_{\rm out})^\dagger (\beta +\hat{b}_{\rm out}) \rangle_c\nonumber \\
&\sim& |\beta|\langle (\hat{b}_{\rm out}e^{i\varphi}+ \hat{b}_{\rm out}^\dagger e^{-i\varphi}) \rangle_c
\ea
where $\mathcal{I}^h(t)$ is the homodyne current as a function of time, $\beta=|\beta|e^{i\varphi}$ is the classical field, and in the second line we have subtracted off the constant term $\sim |\beta|^2$, and dropped lower order terms. As well, the expectation value is defined by
\be
 \langle...  \rangle_c \equiv  \frac{\langle \Psi_c(t)|...|\Psi_c(t)\rangle}{\langle \Psi_c(t)|\Psi_c(t)\rangle},
\ee
where the subscript $c$ refers to the wavefunction {\it conditioned} on the measurement of the outside field. 
Applying the results of input-output theory, we have that the output mode of the cavity $\hat{b}_{\rm out}$ can be given by 
\be
\label{eq:input_output}
\hat{b}_{\rm out} = \hat{b}_{\rm in}+\sqrt{\kappa} \hat{d}
\ee
where $\hat{b}_{\rm in}$ is the cavity input (in this case, just vacuum fluctuations) and $\hat{d}$ the mode inside the cavity. Using this result, and setting $\beta=1$ (w.l.o.g.), we obtain  
\be
\label{eq:derive_homodyne_current}
\mathcal{I}^h(t) =\sqrt{\kappa} \langle(\hat{d} e^{i\varphi}+ \hat{d}^\dagger e^{-i\varphi})\rangle_c + \xi(t).
\ee
Included in this expression is the signal term $\propto\sqrt{\kappa}$, which depends on prior random measurement outcomes (back-action) through the evolution via the SSE, and $\xi(t)$, the shot noise inherited from the vacuum input. A more detailed analysis of how one extracts the signal in practice is given in Appendix~\ref{appendix:noise}.

Given this measurement of the output field, the SSE for the (unnormalized) conditional wavefunction $|\Psi_c(t)\rangle$ of the state of the system is
\ba
\label{eq:SSE_fullsystem}
(I) \, d|\Psi_c(t)\rangle &=& \bigg[\left(-i\hat{H}_{\rm bh}-ig \hat{a}_1^\dagger \hat{a}_2 \hat{d}^\dagger -i g^* \hat{d} \hat{a}_2^\dagger \hat{a}_1 \right)dt \nonumber \\ &&  \big(-\frac{\kappa}{2} \hat{d}^\dagger \hat{d} \big)dt+ \sqrt{\kappa}\,  e^{i\varphi} dq(t) \,\hat{d}\bigg] |\Psi_c(t)\rangle,
\ea
where the state vector $|\Psi_c(t)\rangle$ now acts on the system (plus cavity) alone, as the measurement of the electric field has yielded the eigenvalue $dq(t)$,
\be
dq(t)=\sqrt{\kappa}\langle \hat{d}e^{i\varphi}+\hat{d}^\dagger e^{-i\varphi}\rangle_c dt + dW(t),
\ee
for $dW(t)$ is a random number whose probability density process is that of a Wiener increment. This evolution can equivalently be written in terms of the homodyne current \cite{Carmicheal:1993aa}
\ba
\label{eq:SSE_fullsystem}
(I) \, d|\Psi_c(t)\rangle &=& \bigg[\left(-i\hat{H}_{\rm bh}-ig \hat{a}_1^\dagger \hat{a}_2 \hat{d}^\dagger -i g^* \hat{d} \hat{a}_2^\dagger \hat{a}_1 \right)dt \nonumber \\ && - \big(\frac{\kappa}{2} \hat{d}^\dagger \hat{d} - \sqrt{\kappa}\, \hat{d}  \,\mathcal{I}^h(t)\big) dt\bigg] |\Psi_c(t)\rangle.
\ea
The SSE in Eq.~\eqref{eq:SSE_fullsystem} describes the two sources of evolution: the terms in the first line describe the Hamiltonian evolution, while the terms in the second line describe the impact that recording the measurement $\mathcal{I}^h$ has on the system (back-action).

\subsubsection{Derivation of the Associated Master Equation}
\label{sec:ME_derivation}
While the time evolution of a single run of the experiment, described by the SSE in Eq.~\eqref{eq:SSE_fullsystem}, is conditioned on a given homodyne trajectory $\mathcal{I}^h(t)$, the time evolution of the associated density matrix is described by the average over all possible measurement outcomes. Defined by $\rho(t)={\rm Tr}_B [|\Psi(t)\rangle \langle \Psi(t)|]$, the density matrix evolution can then be derived directly from Eq.~\eqref{eq:Ito_QSSE} to obtain the Master equation
\ba
\label{eq:ME_homodyne}
\frac{d}{dt} \rho(t)&=&-i\left[\hat{H}_{\rm sys},\rho(t)  \right] +\kappa \hat{d} \rho(t) \hat{d}^\dagger \nonumber \\ &&-\frac{\kappa}{2} \hat{d}^\dagger \hat{d} \rho(t)-\frac{\kappa}{2}\rho(t) \hat{d}^\dagger \hat{d} 
\ea
where we have used the following 
\ba
{\rm Tr}_B [ dQ |\Psi(t)\rangle \langle \Psi(t)| dQ] &=&\frac{dt}{2}\rho(t) \nonumber \\
{\rm Tr}_B [ dQ |\Psi(t)\rangle \langle \Psi(t)| ] &=&0.
\ea

\subsubsection{Adiabatic Elimination of the Cavity}
\label{sec:adiabatic_el}
In the previous sections we derived the time evolution of the state vector (described by the SSE) and for the density matrix (described by the Master Equation) for the system plus cavity mode. Because the cavity's role is purely for measurement, here we consider the case where $\kappa$, the damping rate of the cavity, is the dominant time-scale, implying that the cavity follows the system dynamics essentially instantly. This is the so-called bad cavity limit, which allows us to eliminate the cavity and simplify the equations. 

Projecting the QSSE of Eq.~\eqref{eq:Ito_QSSE} state vector on the cavity vacuum ($|\Psi(t)^0\rangle$) and singly excited cavity state ($|\Psi(t)^1\rangle$) we obtain a set of coupled equations of motion. 
If $\kappa$ is the dominant scale, one can eliminate the state $|\Psi(t)^1\rangle$,
\be
\frac{\kappa}{2} dt |\Psi(t)^1\rangle \sim \left[ -ig^* \hat{a}_2^\dagger \hat{a}_1  dt + \sqrt{2\kappa} (d\hat{Q}(t))  \right] |\Psi(t)^0\rangle
\ee
obtaining an independent QSSE for $|\Psi(t)^0\rangle$. The resulting wavefunction $|\psi_c(t)\rangle$ -- now just acting on the system involving $\hat{a}_{1/2}$-- obeys the equation of motion
\ba
\label{eq:final_SSE}
(I) \;d|\psi_c(t)\rangle &=& \left(-i \hat{H}_{\rm sys} - \frac{\gamma}{2} \hat{c}_{a}^\dagger\hat{c}_a  + \sqrt{\gamma}\hat{c}_{a}   \,\mathcal{I}^h(t) \right)dt   |\psi_c(t)\rangle \nonumber \\
\ea
with the identification of the following parameters:
\ba
\label{eq:jump_asym}
g=|g|e^{i\phi,}; \;\gamma =  \frac{4|g|^2}{\kappa}; \;\hat{c}_{a}  = -ie^{i\phi}\hat{a}_1^\dagger \hat{a}_2,
 \ea
 where the we have labelled the jump operator $\hat{c}_{a}$ to distinguish this asymmetric scheme from the symmetric scheme given in the next section. In terms of these parameters, the homodyne current can be written as
\be
\label{eq:homodyne_current}
 \mathcal{I}^h(t) =\sqrt{\gamma} \langle(\hat{c}_{a}e^{i\varphi}+ \hat{c}^\dagger_{a} e^{-i\varphi}) \rangle_c + \xi(t).
\ee
 To relate this final result to the atomic current for the Bose-Hubbard Hamitlonian, let us recall its expression given in Eq.~\eqref{eq:current_example}. From this, we see that if we fix the phase $\varphi$ as
 \be
 \varphi=\theta-\phi
 \ee
 then homodyne current gives the desired form, 
\be
\label{eq:homodyne_current_asym}
 \mathcal{I}^h(t) =\sqrt{\gamma}\langle\hat{\mathcal{J}}_{1,2} \rangle_c + \xi(t).
\ee
The associated Master equation for the system with the cavity eliminated is  
\ba
\label{eq:ME_homodyne}
\frac{d}{dt} \rho(t)&=&-i\left[\hat{H}_{\rm bh},\rho(t)  \right] +\gamma \hat{c}_a \rho(t) \hat{c}_a ^\dagger \nonumber \\ &&-\frac{\gamma}{2} \hat{c}_a ^\dagger\hat{c}_a  \rho(t)-\frac{\gamma}{2}\rho(t) \hat{c}_a ^\dagger \hat{c}_a . 
\ea
\subsubsection{Expanding to Multiple Sites}
\label{sec:multiple_sites}
While this derivation has been done considering a single transition between two states ($\hat{a}_{1/2}$) the derivation can be simply expanded to involve any number of transitions present in the system. Coupling the cavity to multiple transitions will change the operator $\hat{c}_a$ to
\be
\hat{c}_a=-ie^{i\phi}\sum_j\hat{a}_j^\dagger \hat{a}_{j+1}.
\ee
where the sum includes any transitions coupled to the cavity mode. For the form of the current derived from the Bose-Hubbard Hamiltonian, such a coupling results in a homodyne measurement proportional to the sum of the associated local currents
\be
 \mathcal{I}^h(t) =\sqrt{\gamma}\sum_j \langle\hat{\mathcal{J}}_{j,j+1} \rangle_c + \xi(t).
\ee
In particular, if all transitions are coupled to the cavity, the homodyne current reflects the global current $\hat{\mathcal{J}}_{\rm tot}$ of the system. Qualitatively, measuring the global current can lead to completely different dynamics, as the resulting jump operator commutes with the kinetic term in the Hamiltonian $\hat{H}_{\rm bh}$. This implies that the system has the chance to evolve into a joint eigenstate, where as non-commuting operators will always lead to competition in the evolution. This difference behavior while measuring local or global currents will be studied in detail in the next section. 

\begin{figure}[t!]
\begin{center}
	\includegraphics[width=0.45\textwidth]{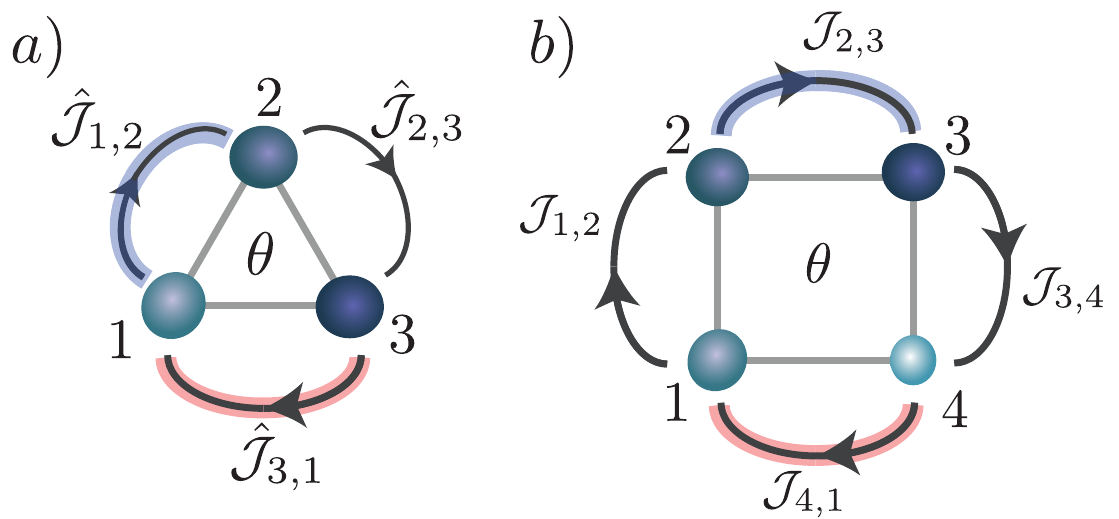}
	\caption{Coupling transitions to different cavities can result in measurement of a) non-commuting operators $\mathcal{J}_{1,2}$, $\mathcal{J}_{2,3}$ or b) commuting operators $\mathcal{J}_{4,1}$, $\mathcal{J}_{2,3}$. Evolution is given by the SSE with multiple couplings and jump operators, as given in Eq.~\eqref{eq:multiple_cavities}. Realisations of such setups could be made in both external (lattice) systems or made up from internal states.}
	\label{fig:triangle_square}
\end{center}
\end{figure}

Instead of coupling different transitions to the same cavity mode, one can also consider coupling different transitions to different cavities, resulting in simultaneous measurements of multiple different local currents. By generalising the above derivation for more than one cavity one obtains a SSE of the form 
\ba
\label{eq:multiple_cavities}
(I) \;d|\psi_c(t)\rangle &=& \bigg[\left(-i \hat{H}_{\rm sys} - \sum_m\frac{\gamma_m}{2} \hat{c}_{a,m}^\dagger\hat{c}_{a,m}  \right)dt  \nonumber \\ && \qquad+  \sum_m\sqrt{\gamma_m}\hat{c}_{a,m}   \,\mathcal{I}^h_m(t) dt \bigg] |\psi_c(t)\rangle 
\ea
where the $m$ now labels each cavity, $\gamma_m, \hat{c}_m$ are the corresponding measurement strength/operator, respecitvely, and $\mathcal{I}^h_m$ is the homodyne current measured from each coupled cavity. In this case, even beyond commuting with the Hamiltonian, the jump operators may not commute with each other, a simple example of which is shown in Fig.~\ref{fig:triangle_square}. For measurements of local current on neighboring sites `share' the middle site, and the operators will not commute (Fig.~\ref{fig:triangle_square} a)) while for measurements of local currents that do not share sites, the operators will commute (Fig.~\ref{fig:triangle_square} b)).

Additionally, we recall that here `lattice sites' can be both sites of a spatial lattice, as in an optical lattice, or also an internal degree of freedom of an atomic cloud (`synthetic dimension'). The strength of the signal, and thus the signal to noise ratio, will depend on the number of atoms at site $j$ and $j+1$. While it may be challenging for occupation with single atoms to measure the current on a single link in a optical lattice ($\hat{\mathcal{J}}_{j,j+1}$), this signal will be strongly enhanced in the presence of many atoms, as in measurement of a global current ($\hat{\mathcal{J}}_{\rm tot}$). This could be even more enhanced when considering the setup of Fig.~\ref{fig:triangle_square} a), where the sites of a triangular lattice represent three internal state BECs with potentially a {\em large number of atoms} coupled by Raman fields. 


Finally, in the case where multiple sites couple to the same cavity field, there will be second order interactions between sites mediated by the cavity field. Its energy scale, second order in the cavity coupling, can be made significantly smaller than the strength of the current measurement by requiring the condition $\Omega^2\gg\Delta\kappa$, for $\Delta$ the detuning to the Raman excited state manifold, which can be achieved in experiment easily. 


\subsection{Symmetric Raman Coupling to the Cavity}
In the previous section, we derived the SSE and the expression for the homodyne current for the first, asymmetric scheme. Here we consider the second microscopic scheme, in which the cavity is coupled {\it symmetrically} to the Bose-Hubbard system.  While in the previous scheme a photon added to the cavity was associated with the transfer of an atom from $\hat{a}_1$ to $\hat{a}_2$ (and {\it not} vice versa), this coupling will result in a photon in the cavity with the transfer of an atom in both directions. Such a coupling would result in the Hamiltonian $H_{\rm sys}$ taking the form
\ba
\label{eq:Hsys_symm}
\hat{H}_{\rm sys}&=& \omega_{\rm cav} \hat{d}^\dagger \hat{d}+\omega_{21}\hat{a}_2^\dagger \hat{a}_2+ \hat{H}_{\rm bh} \nonumber \\ &&
- \left(g_R\hat{d}^\dagger+g_L\hat{d} \right)\hat{a}^\dagger_1\hat{a}_{2}- \hat{a}^\dagger_2\hat{a}_{1}\left(g_L^*\hat{d}^\dagger +g_R^*\hat{d} \right)\nonumber \\ &&
\ea
where $g_{R/L}$ are the effective coupling strengths after the elimination of the excited states, and based on two different Rabi frequencies (see schematic in Fig.~\ref{fig:schematic} b). In order to equally weight each of these processes, we assume $|g_R|=|g_L|\equiv |g|$ such that $g_{R/L}=|g|e^{i\phi_{R/L}}$.

The derivation of the SSE for this scheme is identical to the previous section, simply replacing the Hamiltonian $H_{\rm sys}$ with that of Eq.~\eqref{eq:Hsys_symm}, thus here we just state the results. After elimination of the cavity, the SSE is given by
\ba
\label{eq:final_SSE_symm}
d|\psi_c(t)\rangle &=& \bigg[\left(-i \hat{H}_{\rm sys} - \frac{\gamma}{2} \hat{c}_{s}^\dagger\hat{c}_s  + \sqrt{\gamma}\hat{c}_{s}   \,\mathcal{I}^h(t)   \right)dt \bigg] |\psi_c(t)\rangle \nonumber \\
\ea
with associated jump operator
\be
\label{eq:jump_symm}
\hat{c}_{s} \equiv \left(i e^{i\phi_R} \ha^\dagger_{1}\ha_{2} +i e^{-i\phi_L} \ha^\dagger_{2}\ha_{1} \right),
\ee
and the measurement strength is equivalent to that of the previous section $\gamma=4|g|^2/\kappa$.
For the homodyne current to match the atomic current, the quadrature we measure needs to match that of the atomic current (c.f. Eq.~\eqref{eq:current_example}), thus we want that
\ba
\hat{c}_{s}e^{i\varphi}+ \hat{c}^\dagger_{s} e^{-i\varphi} \propto i e^{i\theta} \ha^\dagger_{1}\ha_{2} -i e^{-i\theta} \ha^\dagger_{2}\ha_{1} 
\ea
This equation can be satisfied if, for example, $\varphi=0; \phi_L=-\phi_R+\pi; \phi_R=\theta$, and one will again obtain the desired expression for the homodyne current
\be
\label{eq:homodyne_current_sym}
 \mathcal{I}^h_{1,2}(t) =\sqrt{\gamma}\langle\hat{\mathcal{J}}_{1,2} \rangle_c + \xi(t).
\ee
The associated master equation takes the same form as the previous scheme and is given by 
\ba
\label{eq:ME_homodyne}
\frac{d}{dt} \rho(t)&=&-i\left[\hat{H}_{\rm bh},\rho(t)  \right] +\gamma \hat{c}_s \rho(t) \hat{c}_s ^\dagger \nonumber \\ &&-\frac{\gamma}{2} \hat{c}_s ^\dagger\hat{c}_s  \rho(t)-\frac{\gamma}{2}\rho(t) \hat{c}_s ^\dagger \hat{c}_s . 
\ea
 Thus, regardless of using the asymmetric or symmetric coupling scheme, one obtains a homodyne current proportional to the atomic current (Eq.~\eqref{eq:homodyne_current_asym} and Eq.~\eqref{eq:homodyne_current_sym}). However, the form of the operator $\hat{c}_{a/s}$ appearing in the SSE (Eq.~\eqref{eq:final_SSE} and Eq.~\eqref{eq:final_SSE_symm}) reflects the asymmetry or symmetry of the microscopic coupling. While this difference seems minimal, in the following sections we will study instances where this can drastically change the physical response of the system.

\section{Atomic Current Measurement in the Bose-Hubbard Model}
\label{sec:BHmodel}
In the last section we concentrated on a single transition between two atomic levels (or external sites) and derived the equations of motion starting from two different microscopic implementations. In this section we will take this building block and expand it to study the evolution of an atomic many-body system with continuous measurement, considering both microscopic implementations. 

The system we will describe is a Bose-Hubbard (BH) system, described by the Hamiltonian
\ba
\label{eq:H_bh_general}
\hat{H}_{BH}&=& -J\sum_{j=1}^L \left( e^{i\theta}\hadag_j\ha_{j+1}+{\rm h.c.} \right)+ U \sum_{j=1}^L\hat{n}_j(\hat{n}_j-1)\nonumber \\
&\equiv&\hat{H}_J +\hat{H}_U 
\ea
where $\ha_j$ refers to the destruction operator at site/state $j=1,\ldots,L$, with corresponding number operator $\hat{n}_j$, $J$ is the hopping coefficient with phase $\theta$, $U$ is the (local) interaction, and we have assumed periodic boundary conditions. 

The phase on the hopping term in Eq.~\eqref{eq:H_bh_general} is imprinted by implementing this term via an external laster, whose phase can be tuned: this is the familiar implementation of synthetic gauge fields for
cold atoms with Raman assisted hopping \cite{Jaksch03}. We emphasize
that in that case the underlying laser beams are described as intense
classical light beams, and thus the phase $\theta$ is a c-number
(parameter), and back-action of atomic motion on the laser beams is
assumed negligible. Instead, in the setup here, the transition includes a coupling to a vacuum-driven cavity mode. As mentioned earlier, the external laser and vacuum-driven cavity mode can be reinterpreted as a coherently-driven cavity, where the coherent part drives the transition and thus imprints the desired
phase (gauge field) onto the hopping matrix element. With our
measurement scheme we are interested here in the back-action on the
light beam itself, which is is detected in a homodyne setup as propotional
to the atomic current (Eq.~\eqref{eq:homodyne_Current}). In this sense, in our CQED setup the gauge
coupling acquires also a dynamical \emph{ }aspect\emph{}. However, we emphasize the conceptual differences to the recent discussion of
implementing of dynamical gauge fields and lattice gauge theories
of high-energy physics model with cold atoms. See Ref.~\cite{Banerjee:2012aa, Wiese:2013aa}.

In the first example below we will consider a BH system with $L=3$ that, in the limit of large particle number, reduces to a two-level system (TLS). While simple, this example can give valuable intuition of the effects of measurement on the system dynamics, and results in the measurement of macroscopic quantum tunnelling between the two states. The second example below will be for the full BH system, where every transition is coupled to the cavity, resulting in measurement of the total atomic current. This, as we will show, is a QND measurement, which implies that the two microscopic implementations result in qualitatively the same time evolution. Finally, the last example is that of a measurement of the local current, which is non-QND. Here the differences between the back-action in the two microscopic coupling schemes will become apparent.  

\subsection{Simple Example: Reduction to a Two-Level System} 
\label{sec:TLS}
Before going to the complete model, the aim of this section is to consider parameters where the system reduces to a two-level system, in order to gain intuition on the dynamics of a many-body system subject to continuous current measurement \footnote{Such a model was also considered in: ``Applications of quantum measurement in single and many body systems", PhD thesis, V. Steixner, (2010) Supervisor: P. Zoller.}. 
We consider a system of $L=3$ sites, and focus on the limit of large total number of particles, $N = \sum_j \langle \hat{n}_j \rangle \gg1$. In this limit, the  Hamiltonian is best written in the phase representation, where the system is described in terms of three phases $\phi_{i}, i=1,2,3$ instead of local densities $n_i$. In this representation, the term in the Hamiltonian $\hat{H}_J$ gives rise to a potential landscape, which is derived in Appendix~\ref{appendix:TLS} and shown in Fig.~\ref{fig:doublewell}.

\begin{figure}[t!]
\begin{center}
	\includegraphics[width=0.45\textwidth]{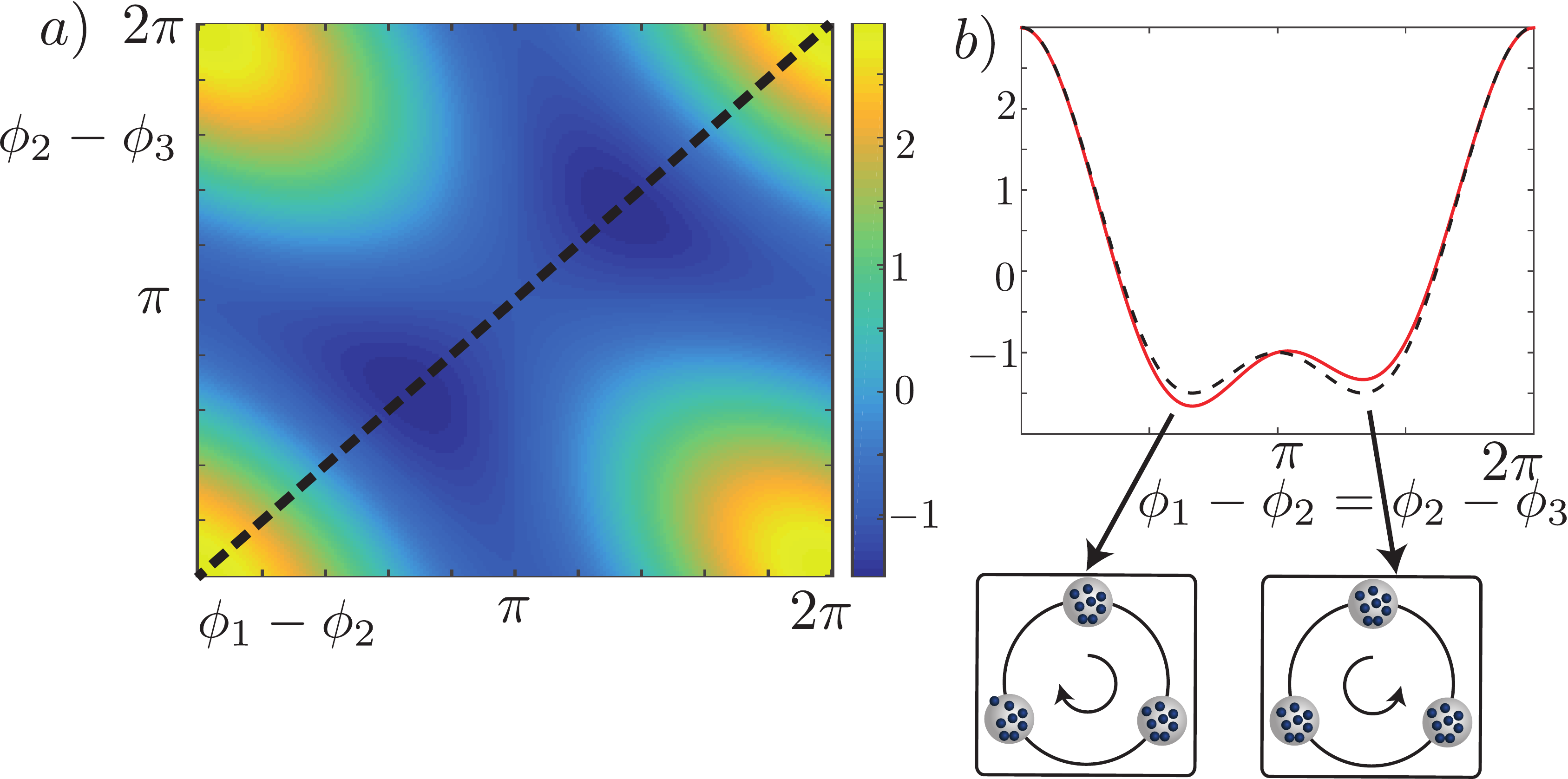}
	\caption{ a) Landscape of the potential term in the phase representation of the BH Hamiltonian (see derivation in Appendix~\ref{appendix:TLS}) with $\theta=\pi$ as function of $\phi_1-\phi_2$ and $\phi_2-\phi_3$. The minima are at $\phi_1-\phi_2=\phi_2-\phi_3= 2\pi/3, 4\pi/3$, and define the two states $|\pm\rangle$ which are characterised by opposite current. b) The double well structure taken from a 1d cut through the potential landscape, indicated on panel a) with a black dashed line. The black/dotted line is for matching parameters to panel a) and the red/solid is for $\theta=1.02 \pi$. In the latter case the double well is offset, into a tilted well.}
	\label{fig:doublewell}
\end{center}
\end{figure} 

From this potential, we can find the ground state structure. In particular, at the point $\theta = \pi$ the potential has double well structure, shown explicitly in Fig.~\ref{fig:doublewell} a). The states of the two wells, which we will label $|\pm\rangle$, have equal and opposite global current,
\be
\langle \pm | \hat{\mathcal{J}}_{\rm tot} | \pm \rangle = \pm \sqrt{3}J N.
\ee
 As one deviates from $\theta=\pi$ the degeneracy of the two wells breaks, and we obtain a tilted-well structure, as shown in Fig.~\ref{fig:doublewell} b). In turn, in the phase representation, the term $\hat{H}_U$ from the Hamiltonian will give rise to the system dynamics. 

In particular, if the well depth is deep enough, the system can be restricted to the Hilbert space containing only the two wells, 
\be
|+\rangle = \begin{pmatrix} 1\\ 0 \end{pmatrix}; \qquad |-\rangle = \begin{pmatrix} 0\\ 1 \end{pmatrix}.
\ee 
Restricted to these two states, the total Hamiltonian $\hat{H}_J+\hat{H}_U$ (See Appendix \ref{appendix:TLS}) can be written as 
\be
\label{eq:TLS_Ham}
\hat{H}_{\rm TLS}= h \hat{\sigma}_z + \omega \hat{\sigma}_x,
\ee
 where $h$ indicates the energy offset of the two wells (set by the deviation of $\theta$ from $\theta=\pi$) and $\omega$ the rate of the tunnelling between the two states (set by the scale of $U/J$). As the two states represent states of equal and opposite global current, the expectation value of the global current operator can be written in terms of the operator $\hat{\sigma}_z$
 \be
\langle \pm | \hat{\mathcal{J}}_{\rm tot} | \pm \rangle =  \sqrt{3}J N \langle \pm | \hat{\sigma}_z | \pm \rangle= \pm \sqrt{3}J N.
\ee

\begin{figure}[t!]
\begin{center}
	\includegraphics[width=0.45\textwidth]{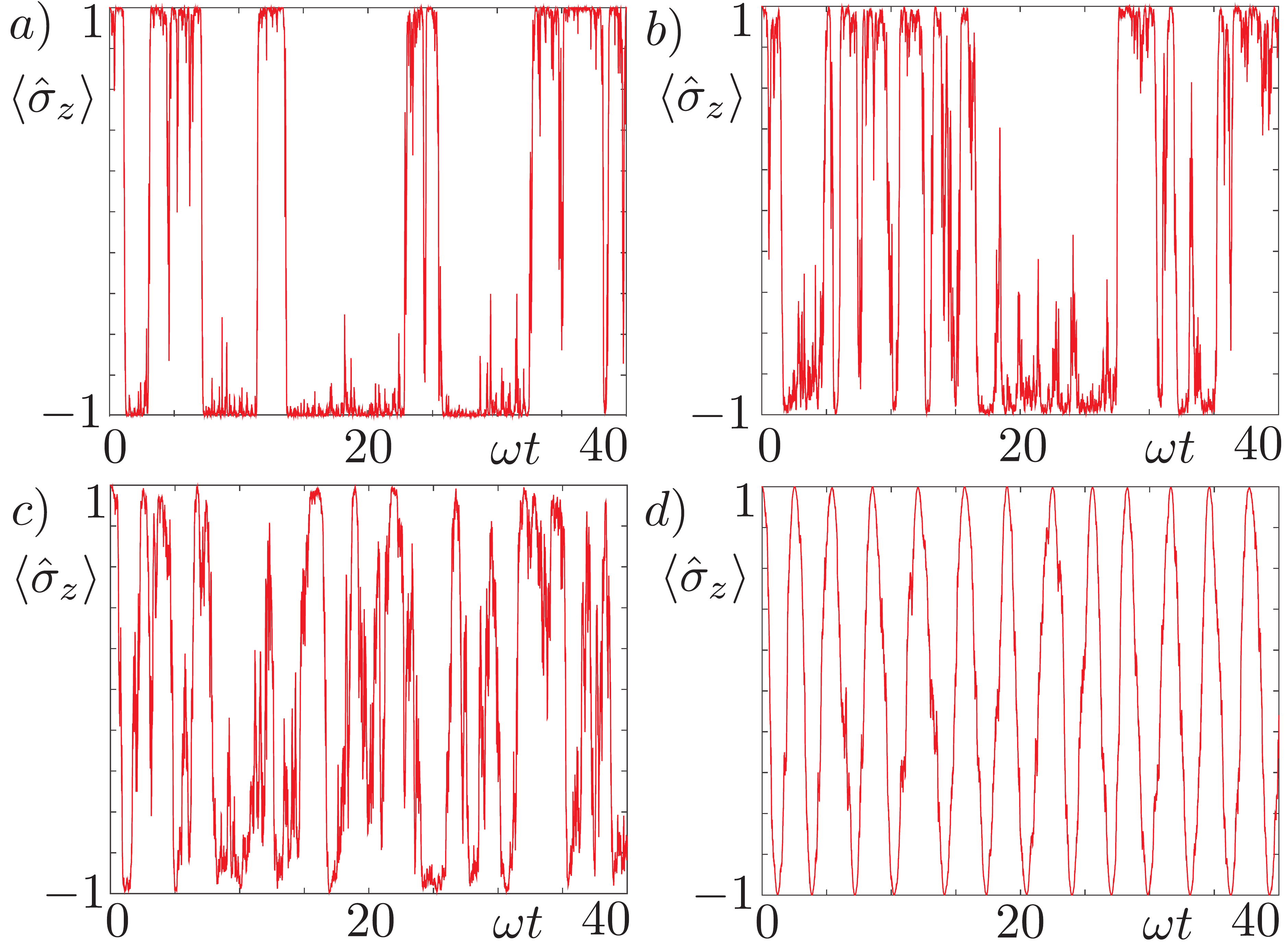}
	\caption{Homodyne Current $\sim \langle \hat{\sigma}_z\rangle$ in a single trajectory of the SSE in Eq.~\ref{eq:SSE_TLS}, with an initial state satisfying $ \langle \hat{\sigma}_z\rangle=+1$. The parameter $\omega$ is the coupling strength between the two states. The ratio $\gamma/\omega$ is $5,2,1,0.5$ from top left to bottom right. For small  $\gamma/\omega$ the evolution is dominated by Rabi oscillations, where the measurement is not strong enough to project the state into one of the two eigenstates. For large $\gamma/\omega$ the measurement acts to project the system onto an eigenstate of $\hat{\sigma}_z$ until tunnelling events, due to finite $\omega$, occur. }
	\label{fig:TLSCurrent}
\end{center}
\end{figure} 

We now want to consider how this system, restricted to the subspace of the two wells, is impacted by the continuous measurement of the global current; ie. what is the impact of the back-action induced when measuring $\langle \pm | \hat{\sigma}_z | \pm \rangle$ continuously in time? To be specific, we consider the asymmetric coupling scheme detailed in Sec.~\ref{sec:measurement_setup} \footnote{As we will see in the next section, in the case of measurement of the global current, using the asymmetric or symmetric scheme will yield qualitatively the same 
dynamics.}, where the hopping term $\hat{a}_j^\dag \hat{a}_{j+1}$ between each adjacent sites is coupled to the same cavity mode, resulting in a homodyne measurement of the global current. Restricting the dynamics to the Hilbert space of the two wells, the SSE governing the system, as derived in Sec.~\ref{sec:measurement_setup}, is 
\begin{equation}
\label{eq:SSE_TLS}
d|\psi_{c}(t)\rangle=\big(-i\hat{H}_{\rm TLS}- \frac{\gamma}{2}\hat{c}_a^\dagger\hat{c}_a+\sqrt{\gamma}\mathcal{I}^{h}(t)\hat{c}_a\big)dt |\psi_{c}(t)\rangle
\end{equation}
where $\hat{H}_{\rm TLS}$ given in Eq.~\eqref{eq:TLS_Ham}, the operator $\hat{c}$ is given by
\be
\hat{c}_a = \frac{1}{2}\hat{\mathbb{1}}+i\frac{\sqrt{3}}{2}\hat{\sigma}_z,
\ee 
and the homodyne current is (up to an overall constant)
\be
\mathcal{I}^h(t)= \sqrt{\gamma} \langle \psi_c(t)|\hat{\sigma}_z |\psi_c(t)\rangle +\xi(t).
\ee
Conditioned on this measurement, we solve the evolution under the SSE in Eq.~\eqref{eq:SSE_TLS} and the results are shown in Fig.~\ref{fig:TLSCurrent}. If the measurement strength is much stronger than the system transition rate (measured by $\omega$, which depends on the rate $U/J$), then the TLS will be pinned to one state, analogous to the quantum Zeno-effect while measuring a state of a qubit. In such a case the homodyne current stays constant, reflecting the current of the pinned state. as $\omega$ increase (by increasing the interaction strength $U$ in the microscopic model), the two states are coupled more and more strongly. This results in macroscopic tunnelling between the two quantum states, indicated by opposite currents. Finally, when the interaction strength dominates the measurement the system continuously transitions between the two eigenstates, analogous to Rabi transitions in a TLS.  The transition between these regimes are shown for single run solutions of the SSE in Fig.~\ref{fig:TLSCurrent}. 

 To emphasize the competition between the coherent Hamiltonian evolution and the measurement back-action, in Fig. 6 we have only plotted the signal part $\propto \gamma$ of the homodyne current, with the shot-noise components excluded. To extract out such a meaningful signal from a measurement record, an appropriate filter should be applied, which depends on the spectrum properties of the signal (see Appendix~\ref{appendix:noise} for details). For weak $\gamma/\omega$, the signal will be mainly Rabi oscillations at frequency $2\sqrt{h^2+\omega^2}$. It can be extracted out by first multiplying $\mathcal{I}^h(t)$ by $\cos(2\sqrt{h^2+\omega^2}t)$ then integration over a finite time $T$, and $\gamma T\gg 1$ would suffice for a large enough signal-to-noise ratio. In the limit of large $\gamma/\omega$, the signal is pinned at $\sigma_z=\pm 1$ due to the quantum Zeno effect, thus direct integration of $\mathcal{I}^h(t)$ over time $T>1/\gamma$ between two neighboring transit events would suffice to manifest such a Zeno dynamics.

\subsection{QND Measurement: Global Current}
\label{sec:QNDMeasurement}
\begin{figure}[t!]
\begin{center}
	\includegraphics[width=0.45\textwidth]{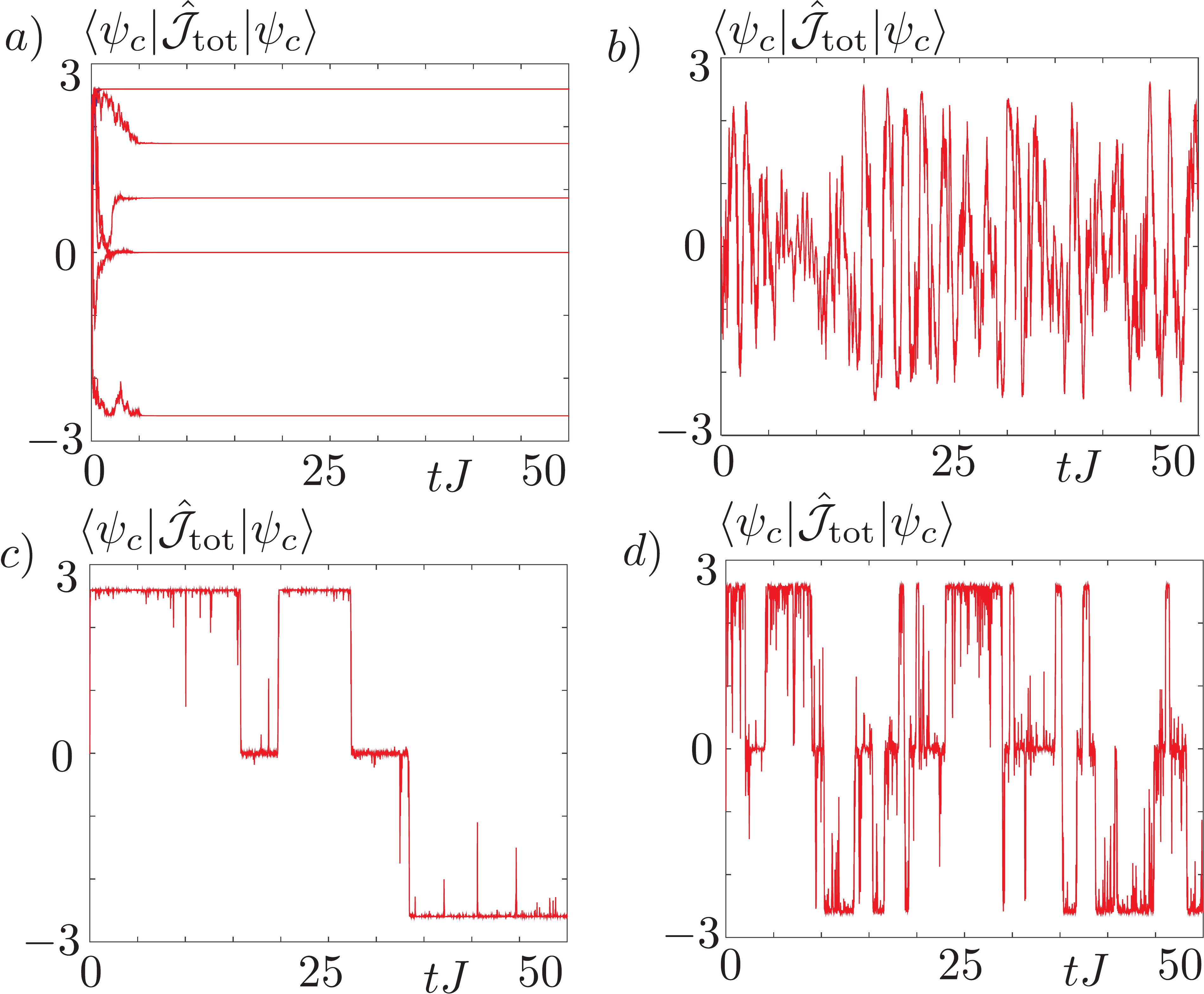}
	\caption{Continuous homodyne measurement of the global current in a system of 3 sites, with 3 particles, at $\theta=\pi/3$. a) 10 different realisations of the SSE at $U=0, \gamma=J$. The measurement is QND and each evolution ends up in one of the eigenstates of the current operator. b)-d) The interplay between the measurement (with strength $\gamma$, which tends to gradually project the system to eigenstates of the global current operator) and interaction (with strength $U$) with respect to the tunnelling rate ($J$); for a weak measurement the state is not completely pinned onto a current eigenstate, as the finite interaction $\propto U$ causes transitions between eigenstates with fixed global current. (b) $U=2J, \gamma=0.1J$). However, for a strong measurement, the state is projected to a current eigenstate, with the interaction causing transitions between these well defined states (c) $U=J/2, \gamma=5J$) and (d) $U=2J, \gamma=5J$).}
	\label{fig:Current}
\end{center}
\end{figure} 

Having gained intuition in the previous section from the simple TLS, in this section we consider the dynamics of the system under the general BH Hamiltonian. We consider coupling each link of the system to the same cavity mode, resulting in a measurement of the global current. This case was introduced briefly in Sec.~\ref{sec:adiabatic_el}, and leads to jump operators of the form 
\ba
\hat{c}_a&=&-ie^{i\phi}\sum_j\hat{a}_j^\dagger \hat{a}_{j+1}, \nonumber \\
\hat{c}_s&=&\sum_j\left(ie^{i\phi_R}\hat{a}_j^\dagger \hat{a}_{j+1}+ie^{i\phi_L}\hat{a}_{j+1}^\dagger \hat{a}_{j}\right),
\ea
for the asymmetric and symmetric couplings, respectively, and by tuning the phases appropriately (see Sec.~\ref{sec:measurement_setup}) the resulting homodyne current will be given by 
\be
\label{eq:homodyne_current}
 \mathcal{I}^h(t) =\sqrt{\gamma}\sum_j \langle\hat{\mathcal{J}}_{j,j+1} \rangle_c + \xi(t).
\ee
In this case, for both schemes the jump operator associated with the measurement commutes with $\hat{H}_J$, i.e.
 \be
 \label{eq:commuting_JumpOPs}
[\hat{c}_s,\hat{H}_J]=0, \; [\hat{c}_a,\hat{H}_J]=0.
\ee
This implies that, for zero interaction ($U=0$ in Eq.~\eqref{eq:H_bh_general}) the measurement is QND for both implementations.  Thus, for each evolution of the SSE, the wavefunction will be projected gradually into one of the joint eigenstates of $\hat{H}_J$ and $\hat{c}_{a/s}$, where the back-action on the system due to the measurement plays no further role. Once in an eigenstate, the homodyne current will simply give the value of the associated current eigenvalue. 

This behavior is shown in Fig.~\ref{fig:Current} a), where we solve the time evolution of a Bose-Hubbard system with 3 sites and 3 particles, at $U=0$. Each realisation begins in a random state, and we see that in each case the system evolves into one of the joint eigenstate of $\hat{H}_J$ and $\hat{c}_{a/s}$, and remains there during the further evolution. This behavior is generic for a BH system with $U=0$, for arbitrary size and particle number.. 

We saw in Section~\ref{sec:measurement_setup} that the difference in the asymmetric and symmetric protocol manifests itself in the form of the back-action on the system. In this QND measurement, the back-action plays no role once the system is projected onto the eigenstates of the total current, thus the long time dynamics of the individual trajectories is qualitatively the same for both the asymmetric and symmetric protocols.

We now consider the effect of a finite interaction. Given that $[\hat{H}_U,\hat{H}_J] \ne 0$, the interaction Hamiltonian does not share an eigenbasis with $\hat{H}_J$ and the measurement operator, and, as expected, will cause transitions between these states. This is shown in Fig.~\ref{fig:Current} b)-d), where we have solved the SSE for a single run at various interaction strengths. If the interaction strength dominates the measurement  (Fig.~\ref{fig:Current} b)) no well defined value of the current is found. As the measurement strength increases, one finds more and more well defined current values, separated by well-defined transitions (Fig.~\ref{fig:Current} d)). As the interaction strength then weakens, the rate of the transitions also decreases (Fig.~\ref{fig:Current} c)). 

Now we briefly discuss how to extract the signal out from the noisy homodyne current $\mathcal{I}^h(t)$. When neglecting the on-site interaction $\hat{H}_U$,  such a QND measurement induces zero back-action to the system. One can thus simply integrate the signal for a time $T\gg1/\gamma$ to reduce the shot noise sufficiently. Including the on-site interaction, the measurement will not be QND and the homodyne current will manifest random transits due to the measurement back-action. As long as $U\ll \gamma$ (cf. Fig. 7 c)), the measurement is nearly QND, there are still long enough time-window between adjacent transits, where one can integrate the signal to achieve a sufficient signal-to-noise ratio. Again, we refer to Appendix~\ref{appendix:noise} for more details on the noise present in the homodyne signal. 

The QND feature of global-current measurement renders individual quantum trajectories qualitatively the same for both the asymmetric-coupling and symmetric-coupling schemes, thanks to the same quantity (namely $\hat{J}_{\rm tot}$) they measure. In contrast, in a non-QND setting, different back-action, depending on the microscopic details of the two coupling scheme, will result in drastically different dynamics of the system. This is illustrated in the next section, where we consider local-current measurement of a BH chain.
\begin{figure}[t!]
\begin{center}
	\includegraphics[width=0.45\textwidth]{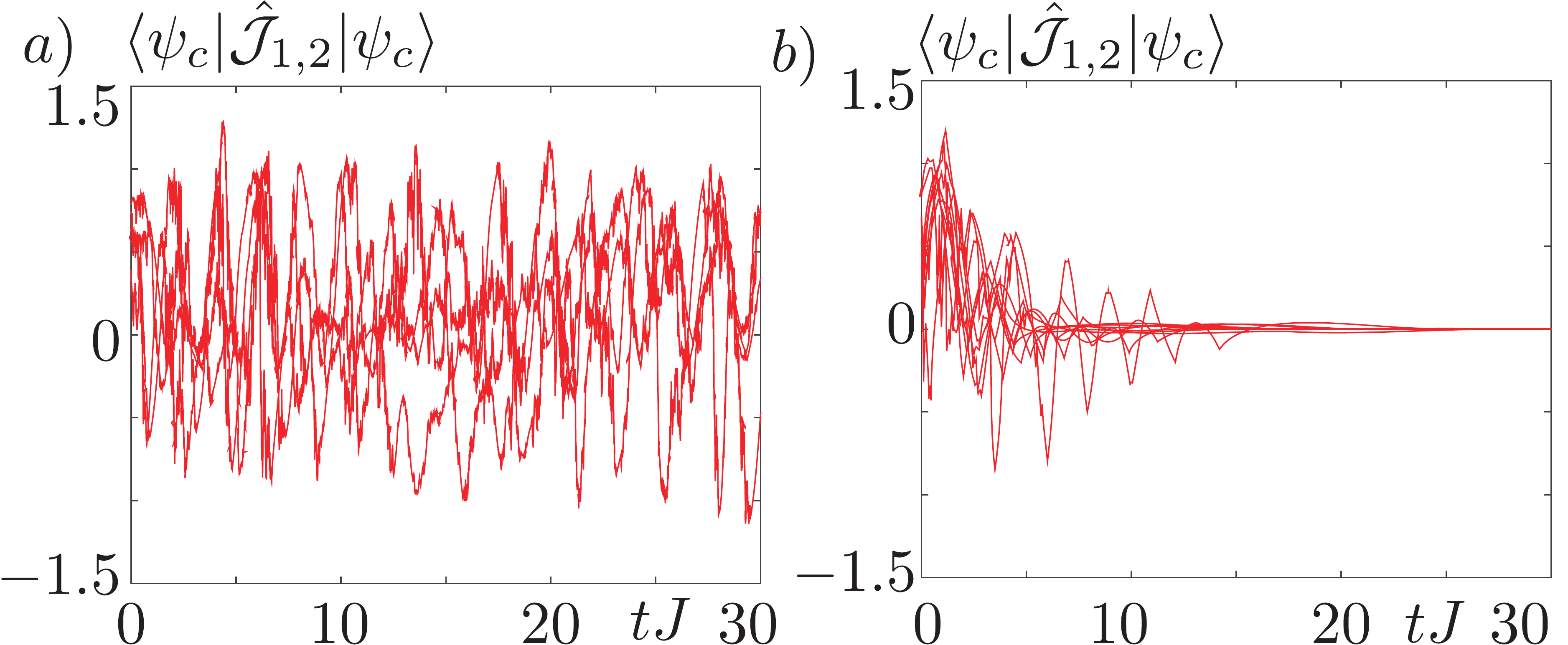}
	\caption{Continuous homodyne measurement of the local current. Plotted are four trajectory solutions to the SSE for a) the symmetric measurement scheme and b) the asymmetric measurement scheme. Parameters, $N=3, L=3, \theta=\pi/3, j=1, \gamma=J, U=0$ are identical for both schemes. In the symmetric case, the measurement is always competing with the Hamiltonian, leading to no well defined current throughout the evolution, whereas in the case of the asymmetric measurement, the system is driven into a dark state (on a time scale determined by $\gamma/J$).}
	\label{fig:localmeasurements}
\end{center}
\end{figure} 
\subsection{Non QND Measurement: Local Current} 
\label{sec:NonQNDMeasurement}
We now turn to the case of measurement of the local current where we will show how the microscopic measurement protocol influences the system dynamics. The local current measurement has associated jump operators defined in Eq.~\eqref{eq:jump_asym} and Eq.~\eqref{eq:jump_symm}, for the asymmetric and symmetric schemes, respectively. In this case neither measurement operator commutes with the Hamiltonian (even at $U=0$)
\be
[\hat{H}_J,\hat{c}_a]\ne 0,\; [\hat{H}_J,\hat{c}_s]\ne 0,
\ee
and thus the measurement is not QND. Therefore, the back-action term of the SSE, proportional to $\hat{c}_{a/s}|\psi_c\rangle$, (see, e.g., Eq.~\eqref{eq:final_SSE}) will always compete with the Hamiltonian evolution. Due to this, the exact form of the measurement operator will determine the back-action response of the system. 

In the symmetric measurement scheme the local current jump operator (Eq.~\eqref{eq:jump_symm}) continuously kicks the system out of the Hamiltonian eigenstates, and when $\gamma \sim J$ the value of the current can never be well defined.  This illustrate the common feature of non-QND measurements - measurement back-action and Hamiltonian dynamics are always competing. Remarkably, in the case of the asymmetric measurement scheme we find that at particular values of $\theta$ the back-action from the measurement drives the system into a dark state satisfying
\be
\label{eq:darkstate}
\hat{c}_a|\psi_{\rm dark}\rangle=0.
\ee
This is illustrated in Fig.~\ref{fig:localmeasurements} b) where the homodyne current is shown for a single realisation of the SSE (with identical parameters as Fig.~\ref{fig:localmeasurements} a)).  Eq.~\eqref{eq:darkstate} holds for $\theta = 2\pi m/L$; commensurate values where the Hamiltonian has a large degree of symmetry and allows for degeneracies in the associated spectrum, as shown in Fig.~\ref{fig:spectrum3levels} a) for $L=3, N=3$. At these values, the dark state, derived in Appendix \ref{sec:derive_darkstate}, is given by
\begin{figure}[t!]
\begin{center}
	\includegraphics[width=0.45\textwidth]{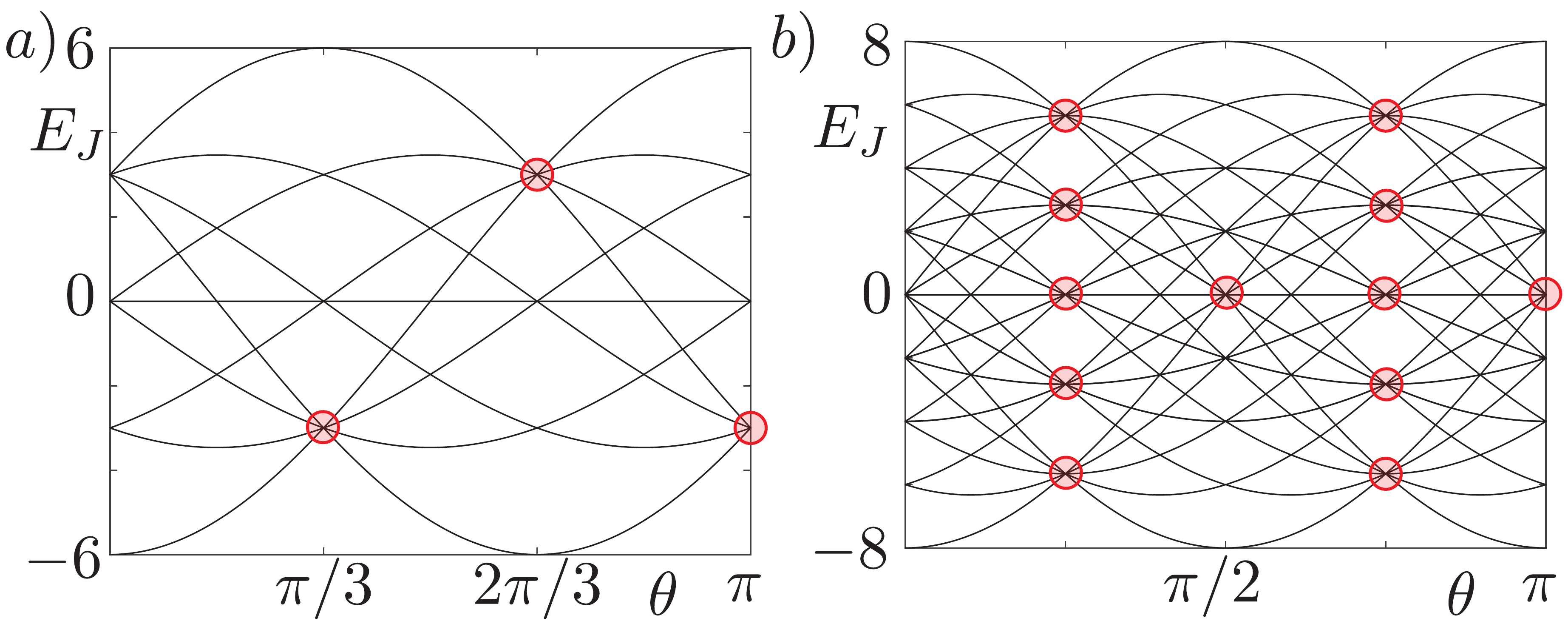}
	\caption{Energy spectrum of the Hamiltonian $\hat{H}_J$ as a function of $\theta$ for a) 3 levels, and $N=3$ and 4) 4 levels and $N=4$. Sets of degenerate states where dark states appear are highlighted in red. For $\theta$ at odd multiples of $\pi/4$ there is a set of dark states, while at even multiples of $\theta=\pi/4$ there is a single dark state.}
	\label{fig:spectrum3levels}
\end{center}
\end{figure} 

 \be
\label{eq:darkstate_realspace}
|\Psi_{\rm dark}\rangle =  \frac{1}{\mathcal{N}} \left(\hat{a}_1^\dagger + e^{i\pi/3}\hat{a}_3^\dagger \right)^N |0\rangle.
\ee

Thus, in response to a purely local measurement, the system is driven into a condensate state, where each of the $N$ atoms occupies the same non-local quantum state. The evolution into this state in single trajectories is shown in Fig.~\ref{fig:localmeasurements} a), where we have solved the SSE for four random initial states. 

In the example of $L=3$, the case with $\theta=2\pi/3$ and $\theta=\pi$ are completely analogous, resulting in the system being driven into a condensate within the set of degenerate states at that point. As well, in this example, the dark state is unique, as there is only one set of degenerate states which is large enough (i.e. has $N+1$ states) to host a dark state (see Appendix \ref{sec:derive_darkstate} for details). However, there are also values of $\theta$, for which there are multiple sets of degeneracies, as exemplified in Fig.~\ref{fig:spectrum3levels} b) for $L=4, N=4$. For each degeneracy point there is one associated dark state, labelled by $k=0,1,...,N$
\ba
\label{eq:fractionalised_condensate}
|\psi_{\rm dark}\rangle_k &=& \frac{1}{\mathcal{N}}\left( \hat{a}_1^\dagger -i\hat{a}_3^\dagger+(1+i)\hat{a}_4^\dagger \right)^k\nonumber \\
&&\;\times \left( \hat{a}_1^\dagger +i\hat{a}_3^\dagger+(1+i)\hat{a}_4^\dagger \right)^{N-k}|0\rangle,
\ea
which, for each $k$ is a fractional condensate state. The SSE will drive the system into a random superposition of these states.

As in the case of the QND measurement, given that $[\hat{H}_U,\hat{H}_J]\ne 0$, when the ratio of parameters $U/J$ and $U/\gamma$ become large enough, the presence of the interaction term will drive the system out of the dark state. 

In a general non-QND measurement setting as exemplified by Fig. 8 a), the signal-to-noise ratio depends on the ratio of $\gamma/J$, which quantifies the measurement strength relative to the energy scale of system dynamics, as well as the width $T$ of the integration time window of the homodyne current $\mathcal{I}^h(t)$. More specifically, we need $JT\ll 1$ to resolve the unconditional dynamics of the system (of which the bandwidth is $~J$) and $\gamma T>1$ to achieve a reasonably good signal-to-noise ratio. The parameters chosen in Fig. 8 a) represents a lower threshold of these conditions. In contrast, the appearance of a pure dark state in the asymmetric driving scheme allows unique possibilities to rescue the signal from the noise sea. Here the measurement back-action is zero and the system is in a well defined dark state, one can simply integrate the signal over a sufficiently long time $\gamma T\gg 1$, to diminish the shot noise.

\begin{figure}[t!]
\begin{center}
	\includegraphics[width=0.35\textwidth]{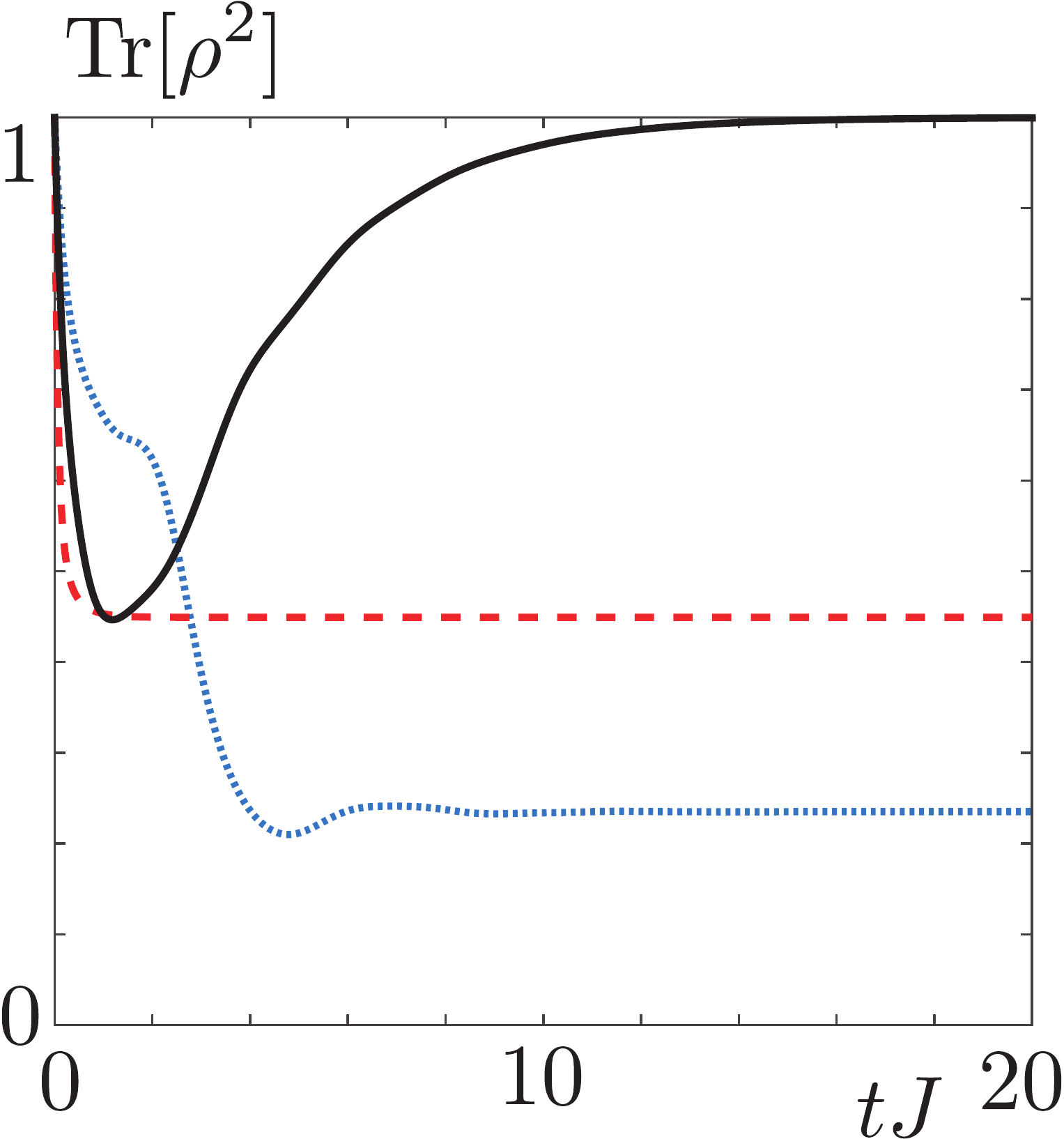}
	\caption{Purity (${\rm Tr}[\rho^2]$) of the density matrix, evolved with respect to the Master Equation. In each evolution a random initial (pure) density matrix is taken. {\bf Red/dashed}: A 3-level system at $\theta = \pi/3$, a QND measurement of the global current $\hat{\mathcal{J}}_{\rm tot}$. Note that the evolution in the case of a QND measurement is initial state dependent (see text for discussion). {\bf Black/solid}: A 3-level system at $\theta=\pi/3$, a measurement of $\hat{\mathcal{J}}_{1,2}$ a non-QND measurement. A dark state exists, and the density matrix evolves into a pure state. {\bf Blue/dotted}: A 3-level system at $\theta=\pi/2$, a measurement of $\hat{\mathcal{J}}_{1,2}$ a non-QND measurement.  No dark state exists, and the system evolves into a totally mixed state for all inital states. }
	\label{fig:darkstates}
\end{center}
\end{figure} 
\subsection{Quantum Reservoir Engineering via Current Coupling}
We now proceed from the single trajectory evolution to consider the evolution of the density matrix with respect to the master equation derived in Sec.~\ref{sec:measurement_setup}. In this case since no measurement outcome is readout, the density matrix evolution describes the driven-dissipative dynamics of the system due to coupling to the measurement apparatus (here the electromagnetic field outside the cavity). Such dynamics have been proposed to be used to engineer desired quantum states, for example, topological states \cite{DIehl:2011, mueller2011:aa, Mueller2011:chapter}. For the BH system considered here, depending on the details of the measurement protocol, the density-matrix evolution manifests variously distinct behaviors, which, in particular, illustrate the heating dynamics of the system. As an example, we consider both the local and global current measurement in the asymmetric-coupling scheme, and illustrate the heating of the system by studying the decrease of purity of its density matrix. The results are shown in Fig. 10, where three distinct heating dynamics are identified, despite that the system is initialized in a (randomly picked) pure state. For local current measurement, for parameters where a dark state exists (black/solid line in Fig.~\ref{fig:darkstates}), the density matrix will evolve into the entangled pure state $\rho(t_f)\rightarrow |\psi_{\rm dark}\rangle\langle \psi_{\rm dark}|$ where $|\psi_{\rm dark}\rangle$ is defined as in Eq.~\eqref{eq:darkstate}. In contrast, for parameters where the dark state condition Eq.~\eqref{eq:darkstate} is no longer satisfied, the density matrix will be driven into a completely mixed state, where the purity will decrease to the lower bound set by the finite-size of the system, as shown in the blue/dotted line of Fig.~\ref{fig:darkstates}. Finally, in the case of a QND measurement (here, measurement of the global current) the steady state of the density matrix will be completely dependent on the initial state. The final purity can range from completely pure, when the initial state is a pure eigenstate, to completely mixed, when the initial state is an equal superposition of all eigenstates of the system. The red/dashed line in Fig.~\ref{fig:darkstates} shows the evolution for an initial state in between these two extremes, where the density matrix maintains finite purity after sufficiently long time.

\section{Outlook}

In conclusion, we have presented a theoretical description of continuous measurement of the atomic current in a cold atom system based on a Cavity QED setup, where the atomic current of atoms moving on a lattice is mapped to a homodyne current. We have illustrated the resulting dynamics for a Bose-Hubbard system, including ring geometries with a synthetic flux, discussing the influence of such factors such as, e.g. if the measurement is QND. In addition we have shown that the microscopic implementation within the CQED setup determines the form of the back-action, and the corresponding master equation were shown to have dark states as steady states. Using this knowledge, we briefly discuss the case of `reservoir engineering', showing how the associated density matrix can be driven into a pure state. 

Continuous observation of a quantum many-body system of cold atoms (see also, e.g., \cite{Molmer:1991aa, Mekhov:2009aa}) should be contrasted to the common measurement scenario of single shot destructive measurements at the end of a given run of the experiment, i.e.~where the preparation and evolution of the quantum system is followed by a (destructive) measurement. An example of the latter is the present implementation of the `quantum gas microscope', allowing for the measurement of local density in both bosonic and fermionic systems to the level of a single site and atom \cite{Bakr2009:aa, Sherson2010:aa, Cheuk2015:aa, Haller2016:aa}. These two ideas, however, exist in a qualitatively different light; While the latter allow the accurate description of the system at a frozen moment in time, the former allows interaction with the system, resulting in the opportunity for specific state preparation or, for example, feedback control, where one can act back on the system based on the measurement outcome.

In a broader context, ongoing research in cold atoms physics has mainly focused on quantum simulation of closed systems, where the goal is to engineer specific quantum many-body Hamiltonians and observe the associated equilibrium and non-equilibrium dynamics, motivated by connections with condensed matter and high-energy physics. In addition, `quantum reservoir engineering' has been introduced, where cold atom systems are coupled in a tailored way to baths,  with the goal of simulating open system dynamics and as a tool for dissipative preparation of interesting quantum states and phases. In contrast, the present work has focused on quantum many-body systems under continuous observation, following the paradigm of continuous measurement as defined and implemented in quantum optics, where single quantum systems are observed in time, and as competition between Hamiltonian dynamics and measurement back-action. 

\section*{Acknowledgments}  
We acknowledge useful discussions with P. Hauke. P.Z. thanks V. Steixner and A. Daley for discussions in the early stages of this work. This work is supported by the Austrian Science Fund SFB FoQuS (FWF Project No. F4016-N23), the European Research Council (ERC) Synergy Grant UQUAM and the EU H2020 FET Proactive project RySQ. The density matrix evolution used in Fig.~\ref{fig:darkstates} was coded in python using the QUTIP library.  
\begin{appendix}
\section{Noise considerations}
\label{appendix:noise}

As discussed in the main text, the measured homodyne current (given in Eq.~\ref{eq:homodyne_current_asym}) includes three parts: First, the signal (i.e. the unconditional expectation value)
\be
s(t) = \sqrt{\gamma} \langle\hat{c}_a e^{i\varphi}+ \hat{c}_a^\dagger e^{-i\varphi}\rangle.
\ee 
Second, the flucturations around $s(t)$ due to the back-action of prior random measurement outcomes (we will define this quantity $\eta(t)$), and thirdly, $\xi(t)$, the shot noise inherited from the vacuum input. The back-action noise manifests in the deviation from the unconditional evolution
\be
\eta(t)\equiv \sqrt{\gamma}\langle\hat{c}_a e^{i\varphi}+ \hat{c}_a^\dagger e^{-i\varphi}\rangle_c  -s(t)
\ee   
and can be quantified by the power spectrum 
\be
S_{\eta \eta}[\omega]=\int dt e^{i\omega t}\langle \eta(t)\eta(0) \rangle
\ee
Similarly, the shot noise can be quantified by its spectrum,
\be
S_{\xi \xi}[\omega]=1
\ee
which takes this simple form for the vacuum noise we consider here. 
In practice, to extract out any meaningful signal from the shot-noise sea, one needs to perform appropriate filtering of the noisy current $\mathcal{I}^h(t)$. The simplest example is to integrate the homodyne current $\mathcal{I}^h(t)$ for a finite time $T$ resulting in a measured homodyne current
\be
\mathcal{I}_T=\int_0^T\;dt\;\mathcal{I}^h(t)
\ee
which selects out its Fourier components within bandwidth $1/T$ around zero frequency. This results in a signal-to-noise given by 
\be
\frac{S}{N}=\frac{T|\bar{s}|^2}{S_{\eta \eta}[0]+S_{\xi \xi}[0]}.
\ee
where $\bar{s}$ is the signal averaged over the time window $T$.
In Sec.~\ref{sec:BHmodel} we have considered several examples of current measurement, and detailed the appropriate filtering of the homodyne current.

\section{Phase representation of the Bose-Hubbard model for $L=3$ and $N\gg1$.}
\label{appendix:TLS}

In Sec.~\ref{sec:TLS} we consider the BH Hamiltonian in the case of $L=3$ sites and with a large number of particles $N\gg1$. In terms of the phase representation the term in the Hamiltonian $\hat{H}_J$ can be written as  
\begin{align}
\label{eq:phi_potential}
&\langle  \phi_1 \phi_2 \phi_3 | \hat{H}_J |\Psi\rangle =\nonumber \\
 & \big( \cos{(\phi_1-\phi_2 +\theta)} +  \cos{(\phi_2-\phi_3+\theta)} + \cos{(\phi_3-\phi_1+\theta)}\big) \nonumber \\
&\qquad \qquad \qquad \qquad \qquad \qquad\times-2J \frac{N}{3}  \langle  \phi_1 \phi_2 \phi_3|\Psi\rangle
\end{align}
where we have introduced the phase representation $|\phi_1 \phi_2 \phi_3\rangle$, satisfying
\ba
\langle \phi_1 \phi_2 \phi_3 |n_1,n_2,n_3\rangle = \frac{e^{i\phi_1 n_1}}{\sqrt{2\pi}} \frac{ e^{i\phi_2 n_2} }{\sqrt{2\pi}} \frac{ e^{i\phi_3 n_3}  }{\sqrt{2\pi}},
\ea
with the conjugate relation $\hat{n}_j =-i \frac{\partial}{\partial\phi_j}$.
Thus, in this representation $\hat{H}_J$ gives rise to the potential landscape, from which we can find the ground state structure. In particular, at the point $\theta = \pi$ the potential has double well structure, shown explicitly in Fig.~\ref{fig:doublewell} a). The states of the two wells, are labeled $|\pm\rangle$, have equal and opposite global current, as given in the main text. 

Similarly, in the phase representation $\hat{H}_U$ can be written as
\ba
\label{eq:HU_phaserep}
\hat{H}_U&=& - U \left(\frac{\partial^2}{\partial\phi_{1,2}^2} +\frac{\partial^2}{\partial\phi_{2,3}^2} +\frac{\partial^2}{\partial\phi_{3,1}^2} \right)+ \frac{UN^2}{3}
\ea
which gives rise to dynamics of the system.

 \section{Dark state derivation}
 \label{sec:derive_darkstate}
 In this section we derive the condition for the dark state evolution, with the goal of understanding why dark states exist in one microscopic scheme and not the other. 
 
\begin{figure}[t!]
\begin{center}
	\includegraphics[width=0.45\textwidth]{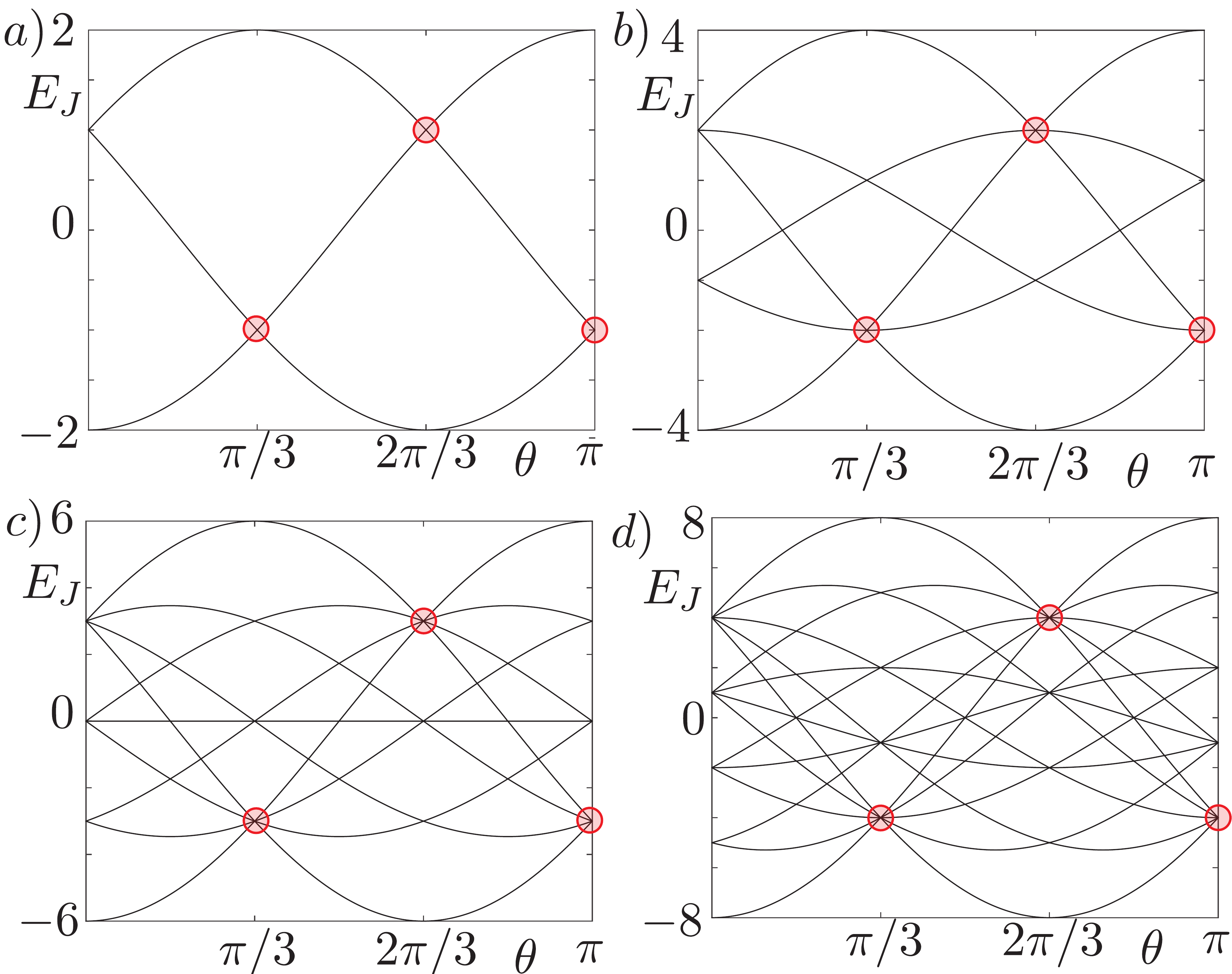}
	\caption{Energy spectrum of the Hamiltonian $\hat{H}_J$ as a function of $\theta$ for a  $L=3$ level system with a different total particle number: a) $N=1$, b) $N=2$, c) $N=3$, d)$N=4$. At the symmetry points, $\theta=\pi/3, 2\pi/3, \pi$ degeneracies emerge, in particular there is one set of $N+1$ degenerate states (highlighted in red). This set of degenerate states will be shown in the text to host a dark state to the local current measurement operator.}
	\label{fig:spectrumdensity}
\end{center}
\end{figure}

To begin, we switch to the momentum basis with operators defined by $\hat{A}_\alpha$, which are the creation operators of the degenerate single-body modes,
\be
\label{eq:momentum_states}
\hat{A}_\alpha^\dagger = \frac{1}{\sqrt{L}} \sum_{k=1}^{L} a_k^\dagger e^{-i\frac{2\pi}{L} \alpha k }
\ee 
where $\alpha=0,\pm1....\pm(L-1)/2$ for $L$ odd, or  $\alpha=0,\pm1....\pm(L-1)/2, L/2,$ for  $L$ even
In terms of these operators the hopping term of the Hamiltonian $\hat{H}_J$ can be written as
\be
\label{eq:MomentumSpaceH}
\hat{H}_J=-2J \sum_\alpha  \cos{\left(\theta - \frac{2 \pi \alpha}{L}\right)} \hat{A}_\alpha^\dagger \hat{A}_\alpha.
\ee
At commensurate values of $\theta$ the Hamiltonian has a high degree of symmetry, allowing for degeneracies in the spectrum. As we will see, these degenerate subspaces in the many-body states will allow the formation of dark states.
In Eq.~\eqref{eq:MomentumSpaceH} we see that the condition for degeneracies in the spectrum is 
\be
 \cos{\left(\theta_{\rm deg} + \frac{2 \pi \alpha}{L}\right)} =  \cos{\left(\theta_{\rm deg} + \frac{2 \pi \alpha'}{L}\right)}
\ee
i.e. when
\be
\theta_{\rm deg}  =  \frac{ \pi (\alpha-\alpha')}{L}.
\ee
At these points, the number of degenerate states depends, of course, on the number of atoms in the system $N$. For example, consider the case where we set $\theta_{\rm deg} = \frac{2\pi (\alpha_1-\alpha_2)}{L}$ for fixed $\alpha_1,\alpha_2$. The single particle Hamiltonian $\hat{H}_J$ is then
\ba
\label{eq:deg_Ham}
\hat{H}_J&=&-2J \sum_{\alpha \ne \alpha_1,\alpha_2}  \cos{\left(\frac{2 \pi (\alpha_1-\alpha_2+\alpha}{L}\right)} \hat{A}_\alpha^\dagger \hat{A}_\alpha \nonumber \\ &&-2J \cos{\left(\frac{2 \pi \alpha_1}{L}\right)} \left( \hat{A}_{\alpha_1}^\dagger \hat{A}_{\alpha_1} +\hat{A}_{\alpha_2}^\dagger \hat{A}_{\alpha_2}  \right)
\ea
This spectrum allows for the buildup of $N$ sets of degenerate states, with the first set having $2$ degenerate states, the second having $3$, and so on, until the last set which has $N+1$ degenerate states. This last set is when all particles are in the degenerate modes, and can be written explicitly as
\be
\label{eq:deg_space1}
|\psi_{\rm deg}\rangle = \frac{1}{\sqrt{x! N-x!} }(\hat{A}_{\alpha_1}^\dagger)^x (\hat{A}_{\alpha_2}^\dagger )^{N-x}|0\rangle
\ee
with $x=0...N$, each having energy 
\be
\langle \psi_{\rm deg}|\hat{H}_J|\psi_{\rm deg}\rangle=-2JN \cos{\left(\frac{2 \pi \alpha_1}{L}\right)}.
\ee
This pattern of degenerate sets of states is shown in Fig.~\ref{fig:spectrumdensity} where we see this set of degenerate states for  $L=3$ and $N=1,2,3,4$. The highest level of degeneracy is always $N+1$, followed by a set of $N$ degenerate states, then $N-1$, ..., and finally the last set has just $2$ degenerate states.

 We note that this discussion assumed that setting $\theta=\theta_{\rm deg}$ resulted only in the degeneracy between the modes $\hat{A}_{\alpha_1}$ and $\hat{A}_{\alpha_2}$. This is not always the case, as for certain values of $L$, there will be a specific $\theta$ which will cause degeneracies simultaneously between different pairs of $\alpha$ values. In particular, this is the case for $L=4$, $\theta=\pi/4$ which will be discussed in in the next subsection. 

Now that we have identified the degenerate subspaces, we show how one can use these degeneracies to construct dark states of the jump operator associate with a local current measurement. 
We begin with the degenerate subspace defined by Eq.~\eqref{eq:deg_space1}, and wish to construct a dark state within this space. To be contained within the degenerate space, the state must have the following form,
\ba
\label{eq:def_darkstate_appendix} 
|\psi_{\rm dark}\rangle &=& \sum_{x=0}^N c_x|\psi_{\rm deg}\rangle \nonumber \\
&=&  \sum_{x=0}^N c_x(\hat{A}_{\alpha_1}^\dagger)^x (\hat{A}_{\alpha_2}^\dagger )^{N-x}|0\rangle
\ea
where, for clarity, in the second line the square root terms ($\sqrt{x!}$ etc...) in the denominator defining $|\psi_{\rm dark}\rangle$ have been absorbed in the constants $c_x$.
Then, in order to be a dark state of the measurement operator, the state must satisfy
\ba
\label{eq:darkstate_condition}
0 &\mathop{=}\limits^! &\hat{c}_a |\psi_{\rm dark}\rangle  \nonumber \\
&\mathop{=}\limits^!& \sum_{\beta,\gamma} e^{2i\pi j \beta/L} e^{-2i\pi (j+1) \gamma/L} A_\beta^\dagger  A_\gamma |\psi_{\rm dark}\rangle
\ea
Taking the form of the dark state defined in Eq.~\eqref{eq:def_darkstate_appendix} this leads to the following set of conditions on the coefficients $c_x$
\ba
\label{eq:simpleconditions}
c_x&=&c_0\prod_{\tilde{x}=1}^x\frac{-(N-\tilde{x}+1)}{\tilde{x}}\left(e^{\frac{-2i\pi \alpha_2 (j+1)}{L}}e^{\frac{2i\pi \alpha_1 (j+1)}{L}}\right)^x \nonumber \\
&=&c_0(-1)^x \frac{N!}{(N-x)!x!}\left(e^{\frac{-2i\pi (\alpha_1-\alpha_2) (j+1)}{L}}\right)^x
\ea
where, to reiterate, $\alpha_1$ and $\alpha_2$ are the two momentum states which have equal energy (see Eq.~\eqref{eq:deg_Ham}), and $j$ labels the dark state with respect to the jump operator $\hat{a}^\dagger_j \hat{a}_{j+1}$. As $x=1...N$, Eq.~\eqref{eq:simpleconditions} represents $N$ conditions which will fix the values of $c_x$ for $x=1...N$. 
The last free parameter $c_0$ is then determined by the condition that 
\be
\langle\psi_{\rm dark}| \psi_{\rm dark}\rangle=1.
\ee 

This set of conditions explains why any degeneracy with less that $N+1$ states will not host a dark state - there are not enough free parameters available to solve
Eq.~\eqref{eq:darkstate_condition}. In particular, this also explains why using the symmetric scheme does not result in a dark state. The expression for a dark state in this case would be 
\ba
0 &\mathop{=}\limits^! &\hat{c}_s |\psi_{\rm dark}\rangle  \nonumber \\
 &\mathop{=}\limits^! &a^\dagger_j a_{j+1} +a^\dagger_{j+1} a_{j} |\psi_{\rm dark}\rangle 
\ea
This condition can never be satisfied within this subspace, thus explaining the difference between these two schemes. 

We now turn to specific examples, and explicitly write down the dark state in these cases.
\subsection{3-level system}

First we consider the case of a 3-level ($L=3$) system. In the momentum basis the Hamiltonian (taking Eq.~\eqref{eq:MomentumSpaceH} for $L=3$) is 
\ba
\hat{H}_J&=&-2J \bigg(  \cos{\theta} \hat{A}_0^\dagger \hat{A}_0+  \cos{(\theta - 2 \pi/3)} \hat{A}_1^\dagger \hat{A}_1\nonumber \\&&+  \cos{(\theta + 2 \pi/3)} \hat{A}_{-1}^\dagger \hat{A}_{-1}\bigg).
\ea
The energy spectrum for the many-body states for $N=1,2,3,4$ as a function of $\theta$ was given in Fig.~\ref{fig:spectrumdensity}, where we see the structure of the energy spectrum described in the previous section: There are three values of $\theta$ where there are degenerate states, occurring at $\theta_{\rm deg} = \pi/3, 2\pi/3, \pi$, consistent with what is shown in Fig.~\ref{fig:spectrumdensity}.
To begin we focus on the point $\theta=\pi/3$. Here the Hamiltonian is 
\ba
\hat{H}_J&=&-J (N-3 \hat{A}_{-1}^\dagger \hat{A}_{-1}\nonumber).
\ea
where we have used that $\hat{A}_0^\dagger \hat{A}_0= N-\hat{A}_1^\dagger \hat{A}_1\nonumber - \hat{A}_{-1}^\dagger \hat{A}_{-1}$. The set of degenerate states from Eq.~\eqref{eq:deg_space1} is
\be
|x\rangle  = \frac{( \hat{A}_0^\dagger)^x( \hat{A}_{1}^\dagger)^{N-x} }{\sqrt{x!N-x!)}}|0\rangle
\ee
where $x=0...N$. These states have energy
\be
\langle x| \hat{H}_J |x\rangle = -JN.
\ee
We will consider the dark state of the operator measuring the local current $\hat{\mathcal{J}}_{1,2}$, between states $j=1$ and $j=2$. Solving Eq.~\eqref{eq:simpleconditions} for the coefficients $c_x$ gives the resulting dark state given in the main text in Eq.~\eqref{eq:darkstate_realspace}.

\subsection{4-level system, $\theta=\pi/4$}
\label{sec:4Lthetapi4} 
We now consider explicitly the case of a 4-level system ($L=4$). Recall in the initial discussion we assumed that setting $\theta=\theta_{\rm deg}$ resulted only in the degeneracy between the modes $\hat{A}_{\alpha_1}$ and $\hat{A}_{\alpha_2}$ (c.f. Eq.~\eqref{eq:deg_Ham}). Here we consider one example where this is not the case. For $L=4$, the Hamiltonian is given by:
\ba
\hat{H}_J&=&-2J  \cos{\theta} (\hat{A}_0^\dagger \hat{A}_0-\hat{A}_2^\dagger \hat{A}_2) \nonumber \\
&&+2J  \sin{\theta} (\hat{A}_{-1}^\dagger \hat{A}_{-1}- \hat{A}_1^\dagger \hat{A}_1).
\ea

We see that at the special point $\theta=\pi/4+m\pi/2, m \in \mathbb{Z}$ the modes labelled by $\alpha=0, -1$  ($\alpha=0, 1$ if $m$ is odd) and the modes $\alpha=1,2$  ($\alpha=-1,2$  if $m$ is odd) are simultaneously degenerate. This results in a higher degree of degeneracy in the spectrum when compared to the previous case of $L=3$. This is shown comparing panels a) to b) in Fig.~\ref{fig:spectrum3levels}. For the case of $\theta=\pi/4$ there are $N+1$ sets of degenerate states, each of which has $N+1$ states. Labelling each set by $k=1...N+1$, the degenerate states belonging to this set can be written as
\be
|x,N-k-x,y,k-y\rangle = \frac{( \hat{A}_0^\dagger)^x( \hat{A}_1^\dagger)^{N-k-x}}{\sqrt{x!(N-k-x!)}}\frac{ ( \hat{A}_2^\dagger)^{y}( \hat{A}_{-1}^\dagger)^{k-y}}{\sqrt{y!(k-y!)}} |0\rangle
\ee
where $x=0...N-k$ and $y=0...k$ with energies
\ba
\langle x,N-k-x,y,k-y| \hat{H}_J|x,N-k-x,y,k-y\rangle =\nonumber \\ -\sqrt{2}J (N-2k) \nonumber \\
\ea
Within each $k$-labelled subspace there is one dark state satisfying $\hat{c}_a|\Psi_{\rm dark}\rangle = 0$, satisfying similar conditions to  and for the local measurement between sites $j=1$ and $j=2$ the dark states are given by satisfying conditions analogous to those in Eq.~\eqref{eq:simpleconditions}. Written in real space, these states are given by Eq.~\eqref{eq:fractionalised_condensate} in the main text. 
These states represent fractional thus the dark state is a fractional condensate state. The SSE will drive the system into a random superposition of these states. However, despite the existence of 5 dark states, none of these dark states are dark for more than one local measurement
\be
\hat{c}_{j,j+1}|\psi_{\rm dark}\rangle_k \ne 0, \; \forall j\ne1 .
\ee
Interestingly, this result means that (only) with a local measurement, do we fix phase coherence in the complete system. 
\section{Effect of Spontaneous Emission}
\begin{figure}[t!]
\begin{center}
	\includegraphics[width=0.45\textwidth]{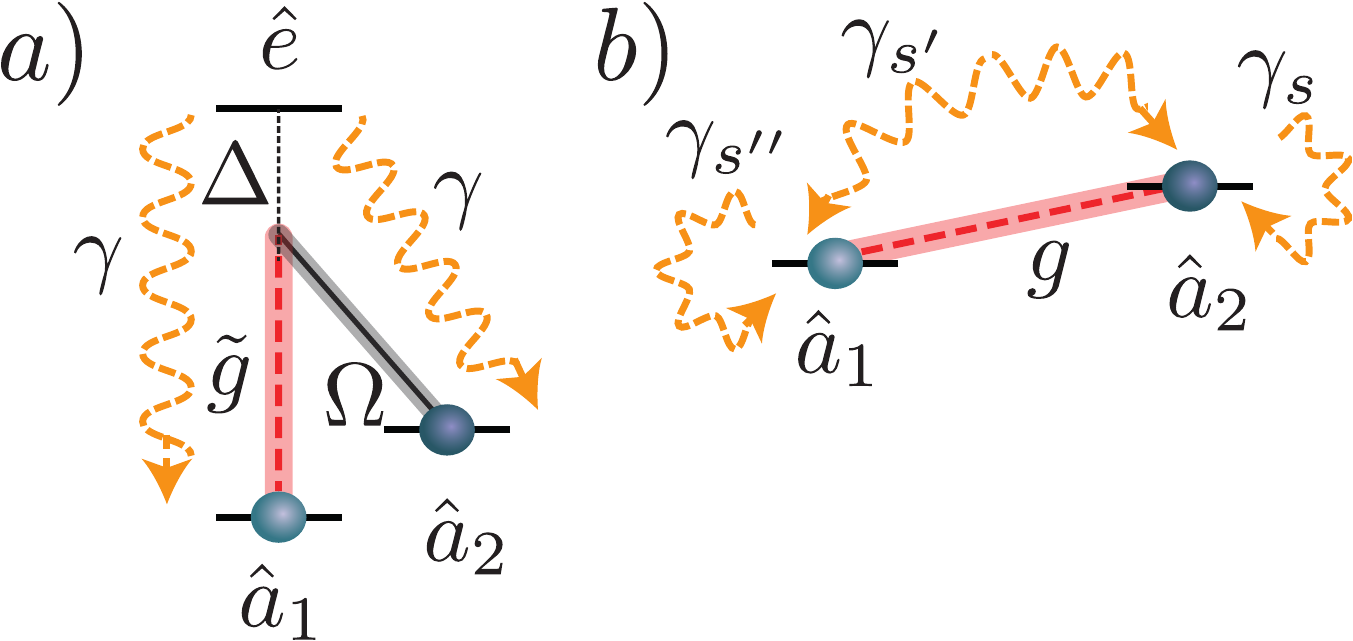}
	\caption{a) Three-level Raman system with coupling to the cavity $\tilde{g}$, and laser driven transition $\Omega$ and detuning to the excited state $\Delta$. We consider spontaneous emission from the excited state to the ground states with equal rate $\gamma$. b) Effective two-level system after elimination of the excited state, resulting in an effective coupling $g$ between the two states, and effective dissipation coupling the two levels $\gamma_s$ and dephasing $\gamma_{s'}$ and $\gamma_{s''}$.}
	\label{fig:3Level_system}
\end{center}
\end{figure} 
In this section we make an analysis of the impact of spontaneous emission on the results presented in the main text. 
We consider spontaneous emission with rate $\gamma$ from the excited state of the Raman transition into both of the ground states, as shown schematically in Fig.~\ref{fig:3Level_system} a). When eliminating the excited state (valid for large detunings) the spontaneous emission from the excited state leads to an effective decay from one ground state to the other, as well as an effective dephasing as illustrated schematically in Fig.~\ref{fig:3Level_system} b). The effective decay rates $\gamma_s$, in the limit of $\Delta \gg \Omega, \tilde{g}$, are given by \cite{Diss2012}
\be
\gamma_s = \left(\frac{\Omega}{\Delta} \right)^2 \gamma,
\ee
where the effective decay rates of the other channels $\gamma_{s'}$ and $\gamma_{s''}$ have similar expressions. 

These photons are emitted into free space (and are not measured, as are the photons which are emitted into the cavity). Thus the evolution of the system can be described by a stochastic master equation, where one decay channel (governed by the decay rate $\gamma_m$) is the measured field leaving the cavity, and where the other decay channels (described by the effect spontaneous emission rates $\gamma_s$) act as standard dissipation processes. The stochastic master equation is 
\ba
\label{eq:SME}
d \hat{\rho}_c(t) &=& -i[\hat{H}_{\rm atom},\hat{\rho}_c]dt\nonumber \\ &&+\frac{\gamma_m}{2}\left(2\hat{c}_m\hat{\rho}_c\hat{c}_m^\dagger - \hat{c}_m^\dagger\hat{c}_m\hat{\rho}_c-\hat{\rho}_c\hat{c}_m^\dagger\hat{c}_m\right)dt \nonumber \\ 
&&+\sqrt{\gamma_m}\left[\left(\hat{c}_m -\langle \hat{c}_m \rangle_c \right) \hat{\rho}_c  +\hat{\rho}_c \left(\hat{c}_m^\dagger -\langle \hat{c}^\dagger_m \rangle_c \right)\right] dW \nonumber\\&&+ 
\sum_s \frac{\gamma_s}{2}\left(2\hat{c}_s\hat{\rho}_c\hat{c}_s^\dagger - \hat{c}_s^\dagger\hat{c}_s\hat{\rho}_c-\hat{\rho}_c\hat{c}_s^\dagger\hat{c}_s\right)dt
\ea
where the jump operators associated with the decay rates $\gamma_s$, $\hat{c}_s$, describe the particular channel (ie. either dephasing $\hat{c} = \hat{a}_j^\dagger \hat{a}_j$ or decay  $\hat{c} = \hat{a}_j^\dagger \hat{a}_{j+1}$).
\begin{figure}[t!]
\begin{center}
	\includegraphics[width=0.35\textwidth]{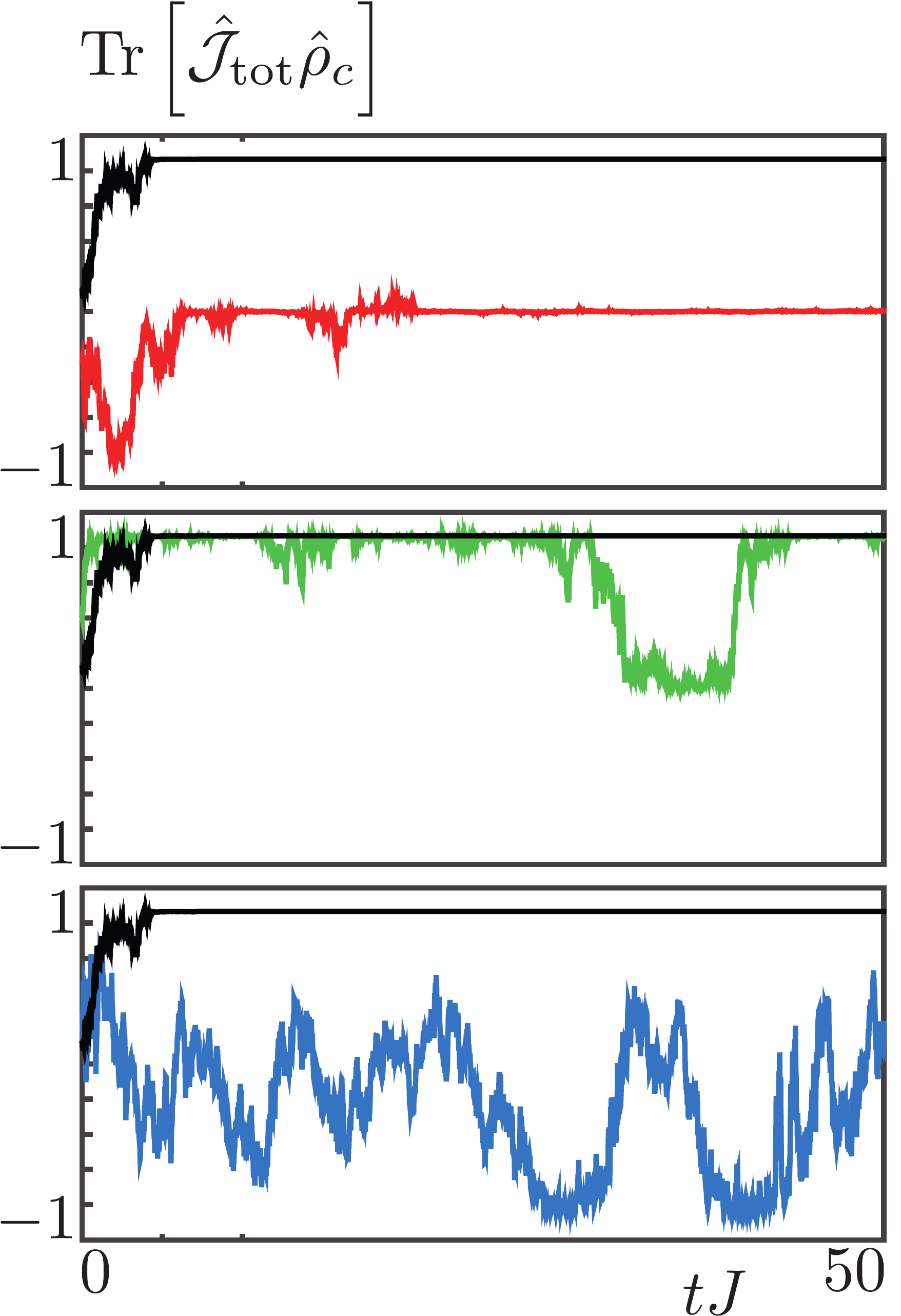}
	\caption{Evolution of ${\rm Tr}[ \hat{\rho}_c  \mathcal{J}_{\rm tot}]$ with respect to the stochastic master equation with one measurement channel and the dissipation channels corresponding to the processes of Fig.~\ref{fig:3Level_system}. Each trajectory is made with a random initial (pure) state. In these trajectories $\theta=\pi$, and all dissipation channels have equal strength $\gamma_{s}= \gamma_{s'}$. Different panels indicate different levels of spontaneous emission: Upper Panel (red) $\gamma_s/\gamma_m=0.05$; Middle panel (green) $\gamma_s/\gamma_m=0.1$; Lower panel
(blue) $\gamma_s/\gamma_m=0.5$. Black in each panel indicates reference for $\gamma_s/\gamma_m=0$.}
	\label{fig:SE}
\end{center}
\end{figure} 
 There are two consequences of including such terms in the evolution: First, additional (non-measured) dissipation channels act as an inefficient detector, in that some photons will leave the system without being detected. Second, if dissipation terms are strong enough they may cause qualitatively different behavior - if the associated jump operators do not commute, they can cause the understanding of the QND measurement, or the evolution into a dark state to break down, as we will explore below. 

\subsection{Impact on Total Current Measurement}
The first example we will consider is the evolution of the system with a total current measurement; a QND measurement. In the analysis in Sec.~\ref{sec:QNDMeasurement}, we found that every state would be driven into a joint eigenstate, which is a steady state of the SSE (c.f. Fig.~\ref{fig:Current}). In Fig.~\ref{fig:SE} we consider this evolution with the additional dissipation channels, but solving the stochastic master equation in Eq.~\ref{eq:SME}.
From this figure we see how the additional processes contradict the QND interpretation of the measurement, when they become on the same order of the measurement. However, for small dissipation rates (in Fig.~\ref{fig:SE}, the red line, with $\gamma_s=0.05\gamma_m$, and even the green line, $\gamma_s=0.1\gamma_m$, the qualitative behavior of the QND measurement can still be seen. 

\begin{figure}[t!]
\begin{center}
	\includegraphics[width=0.35\textwidth]{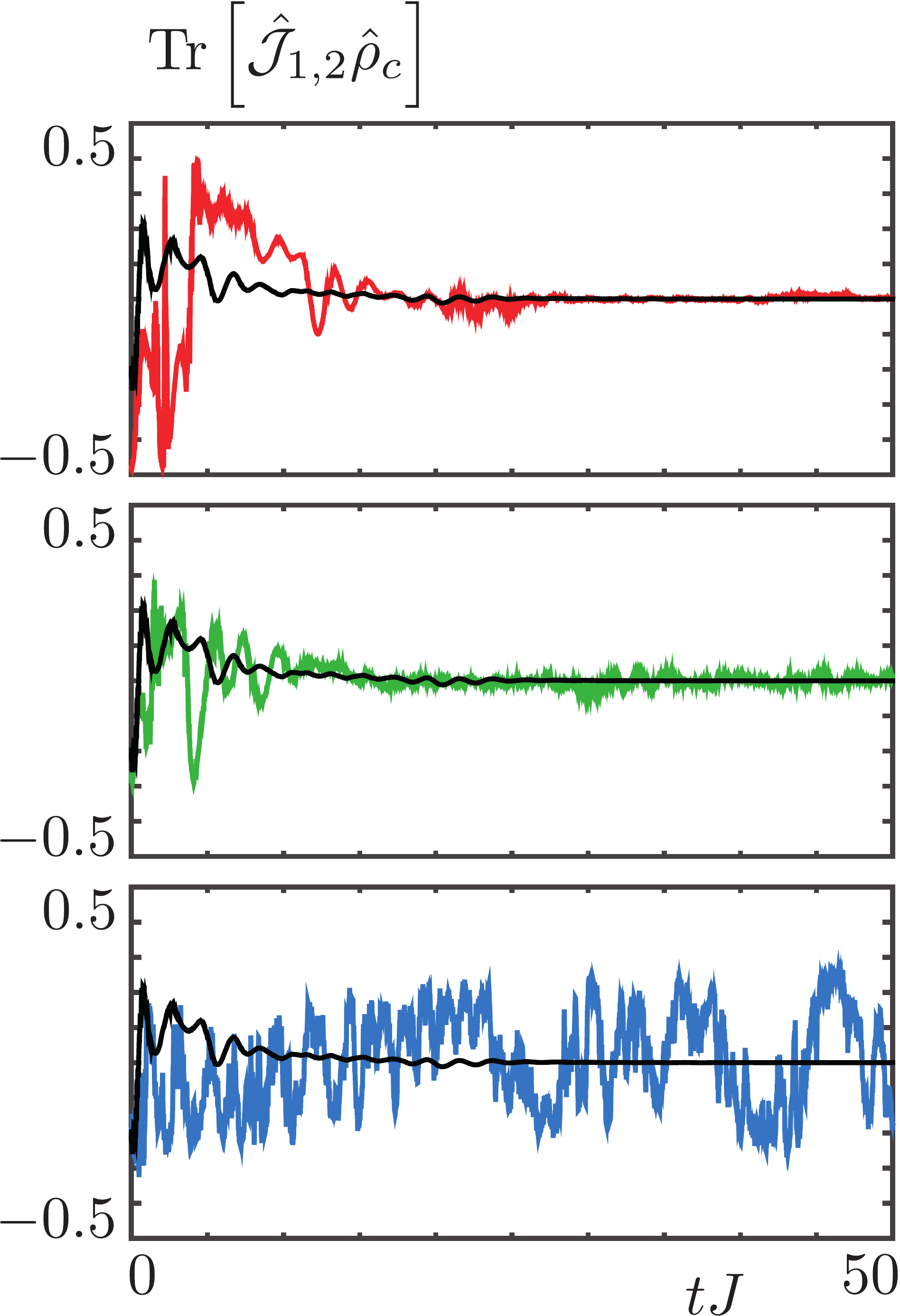}
	\caption{Evolution of ${\rm Tr}[ \hat{\rho}_c  \mathcal{J}_{1,3}]$ with respect to the stochastic master equation with one measurement channel and the dissipation channels corresponding to the processes of Fig.~\ref{fig:3Level_system}. In these trajectories $\theta=\pi$, and all dissipation channels have equal strength $\gamma_{s}= \gamma_{s'}$.  Different panels indicate different levels of spontaneous emission: Upper Panel (red) $\gamma_s/\gamma_m=0.05$; Middle panel (green) $\gamma_s/\gamma_m=0.1$; Lower panel (blue) $\gamma_s/\gamma_m=0.5$. Black in each panel indicates reference for $\gamma_s/\gamma_m=0$.}
	\label{fig:SE_local}
\end{center}
\end{figure} 
\subsection{Impact on Local Current Measurement}

We now repeat the analysis for the case of local current measurement. In the study neglectivg spontaneous emission, we found that at symmetry points, the state would evolve into a dark state (c.f. Fig.~\ref{fig:localmeasurements} in the main text). In Fig.~\ref{fig:SE_local} we show the evolution of the stochastic master equation for the local current measurements in the presence of the dissipation channels introduced above.

Again, similar to the QND case, dissipation at the level of $\gamma_s=0.1\gamma_m$ one can still see the dark state behavior, however at the value of $\gamma_s=0.5\gamma_m$, the dissipation channels are strong enough to drive the system away from the dark state. 
 
\end{appendix}
\bibliographystyle{apsrev}
\bibliography{CurrentRefs}
\end{document}